\documentclass[lettersize, journal, twoside]{IEEEtran} % MODIFIED: added twoside to arguments, by default it assumed oneside
\usepackage{amsmath,amsfonts}
\usepackage{algorithmic}
\usepackage{algorithm}
\usepackage{array}
\usepackage[caption=false,font=normalsize,labelfont=sf,textfont=sf]{subfig}
\usepackage{textcomp}
\usepackage{stfloats}
\usepackage{url}
\usepackage{verbatim}
\usepackage{graphicx}
\usepackage{cite}
\usepackage{cancel}
\usepackage{xcolor}
\usepackage{nicefrac}
\usepackage[inline]{enumitem}
\usepackage{bm}
\usepackage{booktabs}
\usepackage{threeparttable}
\usepackage{tikz}
\usetikzlibrary{positioning}

\newcommand\arxivcopyrighttext{%
	\footnotesize \textcopyright 2026 IEEE.  Personal use of this material is permitted.  Permission from IEEE must be obtained for all other uses, in any current or future media, including reprinting/republishing this material for advertising or promotional purposes, creating new collective works, for resale or redistribution to servers or lists, or reuse of any copyrighted component of this work in other works.}

\newcommand\arxivcopyrightnotice{%
\IEEEpubid{%
	\parbox{\textwidth}{%
		\vspace*{14pt}%
		\centering
		\fbox{\parbox{\dimexpr0.85\textwidth-2\fboxsep-2\fboxrule\relax}{%
				\arxivcopyrighttext}}}}
}

\newcommand{\aligncimrefs}{fpaligncim, digfpcim, fphybrid, intensive-cim-sparse-digital}
\newcommand{\cimrefs}{fpaligncim, digfpcim, fphybrid, intensive-cim-sparse-digital, tsmccim22nm, flexdcim, c2ccim, flexcim, fpcim, oacim, addcim, digfpcim2}

\newcommand{\nbhyphen}{\mbox{-}}
\newcommand{\cu}{C_{\mathrm{u}}}
\newcommand{\circled}[1]{\smash{\tikz[baseline=(char.base)]{
			\node[shape=circle,draw,inner sep=0.5pt] (char) {\small #1};}}}
	
\begin{document}

\title{\fontsize{20}{24}\selectfont Investigating Energy Bounds of Analog Compute-in-Memory\\ with Local Normalization}

% HEADERS FOR ARXIV
% https://journals.ieeeauthorcenter.ieee.org/become-an-ieee-journal-author/publishing-ethics/guidelines-and-policies/post-publication-policies/
% BEGIN
\author{Brian Rojkov, Shubham Ranjan, Derek Wright, Manoj Sachdev, ~\IEEEmembership{Fellow,~IEEE}
\thanks{Corresponding author: B. Rojkov (e-mail: brojkov@uwaterloo.ca)}%
\thanks{The authors are with the Department of Electrical and Computer Engineering, University of Waterloo, Waterloo, ON, Canada.}
\thanks{This work was supported by the Natural Sciences and Engineering Research Council of Canada (NSERC) under Discovery Grant RGPIN-2023-04151.}
}%
\arxivcopyrightnotice
% END

\maketitle
\begin{abstract}
Modern edge AI workloads demand maximum energy efficiency, motivating the pursuit of analog Compute-in-Memory (CIM) architectures. Simultaneously, the popularity of large language models (LLMs) drives the adoption of low-bit floating-point formats that prioritize dynamic range. However, the conventional direct-accumulation CIM accommodates floating-point values by normalizing them to a shared, widened fixed-point scale. Consequently, hardware resolution is dictated by the input's dynamic range rather than its precision, and energy consumption is dominated by the ADC. We address this limitation by introducing local normalization for each input, weight, and multiply-accumulate (MAC) output via a \textit{Gain-Ranging MAC (GR-MAC)}. Normalization overhead is handled by low-power digital logic, enabling the computationally expensive MAC operation to remain in the energy-efficient low-precision analog regime. Energy modelling shows that adding a gain-ranging stage to the MAC increases the input dynamic range by 4 bits without increasing energy consumption at a 35 dB SQNR standard. Additionally, the ADC resolution requirement becomes invariant to input distribution assumptions, allowing the construction of an upper bound with a 1.5-bit reduction compared to the conventional lower bound. These results establish a pathway to unlocking favourable energy scaling trends in analog CIM for modern AI workloads.
\end{abstract}


\section{Introduction}
\begin{figure}[!t]
	\centering
	\includegraphics[width=3.4in]{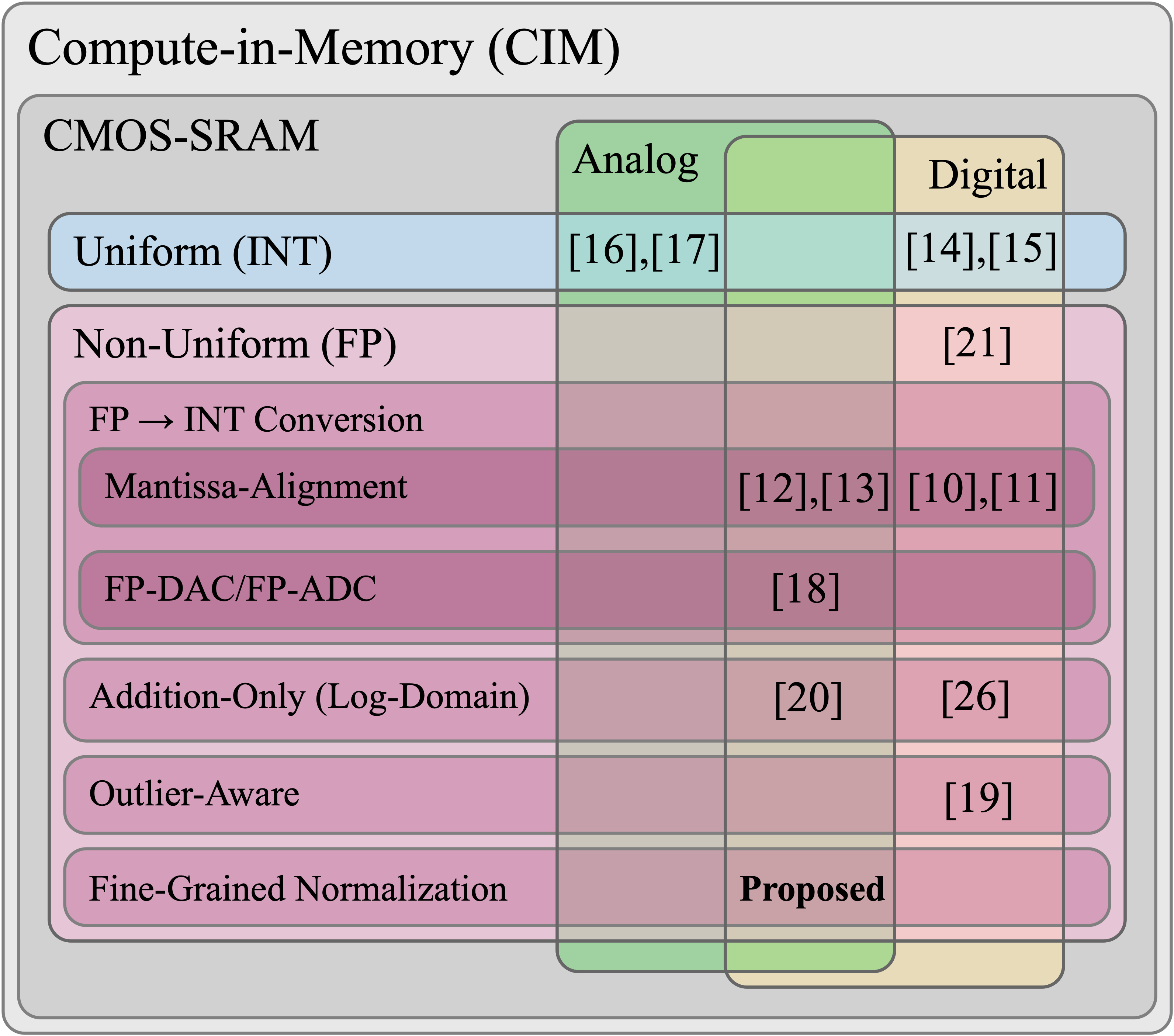}
	\caption{Classification of CMOS-SRAM-based CIMs based on compute domain and quantization type. While emerging Non-Uniform/Floating-Point (\textit{pink}) works offer superior information efficiency over Uniform/Integer (\textit{blue}), prior implementations suffer from increased logic complexity, approximation errors, or overheads introduced by conversion and long-tailed data distributions. Our work (\textit{bottom}) introduces a direct, native floating-point MAC to eliminate these bottlenecks.}
	\label{fig:taxo}
\end{figure}

\IEEEPARstart{C}{ompute-in-Memory (CIM)} architectures lead in energy efficiency figures for deep neural network (DNN) inference workloads, which increasingly dominate modern compute consumption \cite{imc-background}. Research in the CIM paradigm has spanned the architectural \cite{cim-early-arch, cim-early-arch-2}, circuit \cite{cim-early-circuit, cim-early-circuit-2}, and memory-device \cite{cim-early-device, cim-early-device-2, cim-early-device-4, cim-early-device-5} domains. Nevertheless, practical implementations have largely converged on CMOS SRAM-based CIMs \cite{\cimrefs}, owing to their combination of reliability, manufacturability, and favourable power-performance-area (PPA). The taxonomy in Fig. \ref{fig:taxo} categorizes prior art by computation domain and quantization type.

Conventional integer CIM (INT-CIM) works (Fig. \ref{fig:taxo}, \textit{Uniform}) attain high energy efficiency by performing massive in-situ parallel computation, using techniques such as digital bit-serial adder trees to locally accumulate partial products \cite{tsmccim22nm, flexdcim}, or analog charge-domain accumulation one-shot physical dot products \cite{c2ccim, flexcim}. On the other hand, quantization literature has long recognized the information-efficiency advantage offered by floating-point formats, as independent control of range and precision allows the allocation of information across the format's range to better match the distributions seen in neural network weights and activations \cite{why-fp, why-fp-2}. However, energy efficient floating-point CIM (FP-CIM) (Fig. \ref{fig:taxo}, \textit{Non-Uniform}) design remains a challenge, as the exponent breaks the inherent bit alignment structure that underpins the INT\nbhyphen CIM's parallelism. 

\IEEEpubidadjcol
The prevailing solution, \textit{mantissa alignment}, forces compatibility by shifting mantissas by their exponents, effectively converting the data into a widened integer format suitable for conventional CIM arrays \cite{\aligncimrefs}. Unfortunately, this pre-processing incurs an energy overhead and generates long-tailed data distributions with high bit sparsity, forcing the downstream INT-CIM array to overprovision for worst-case dynamic range, thereby wasting energy on redundant precision.

We propose adding a gain-ranging stage to the MAC to enable direct processing of normalized mantissas and to provide a mechanism to reintroduce the exponent's scaling during analog accumulation. Our \textit{Gain-Ranging MAC (GR-MAC)} provides the following benefits:
\begin{enumerate}
	\item \textbf{Relaxed ADC Resolution:} Whereas the conventional CIM ADC resolution requirement depends on input data distribution assumptions, the proposed fine-grained normalization enables a data-invariant upper-bound on the required ADC resolution, reducing it by at least 1.5~bits, and over 6~bits under the empirically observed challenging conditions seen in LLMs.
	\item \textbf{Favourable Energy Scaling Trends:} Increasing the useful input dynamic range of a conventional CIM requires a corresponding increase in the precision (input-referred signal-to-noise ratio) specification. The proposed technique decouples these metrics, enabling an increased input dynamic range without the energy cost of higher-precision data converters.
\end{enumerate}

The benefit of low-bit floating point-formats for LLM inference is that they quantize high dynamic range data without the overhead of high precision. The objective of this work is to provide a path to relay this benefit to CIM hardware. We quantify this by specifying the conventional and proposed CIMs along independent dynamic range and precision axes, and evaluating energy trends across this space in a design exploration study. 

The rest of this paper is organized as follows: Section  \ref{sec:back} reviews existing integer and floating-point CIM implementations. Section \ref{sec:arch} proposes a mixed-signal FP-CIM architecture as a platform for design-space exploration. Section \ref{sec:anal} presents an analysis and discussion based on energy modelling. Finally, Section \ref{sec:conclusion} concludes this work.

\section{Background}\label{sec:back}
Fig. \ref{fig:taxo} provides an overview of the CIM landscape reviewed in this section, distinguishing between the compute domain, quantization type, and FP-CIM implementation strategies. In this section, we will first review the operating principles and scaling limitations of INT-CIM architectures. Next, state-of-the-art FP-CIM architectures are reviewed to identify their specific merits and drawbacks.

\begin{figure*}[!t]
	\centering
	\includegraphics[width=7.16in]{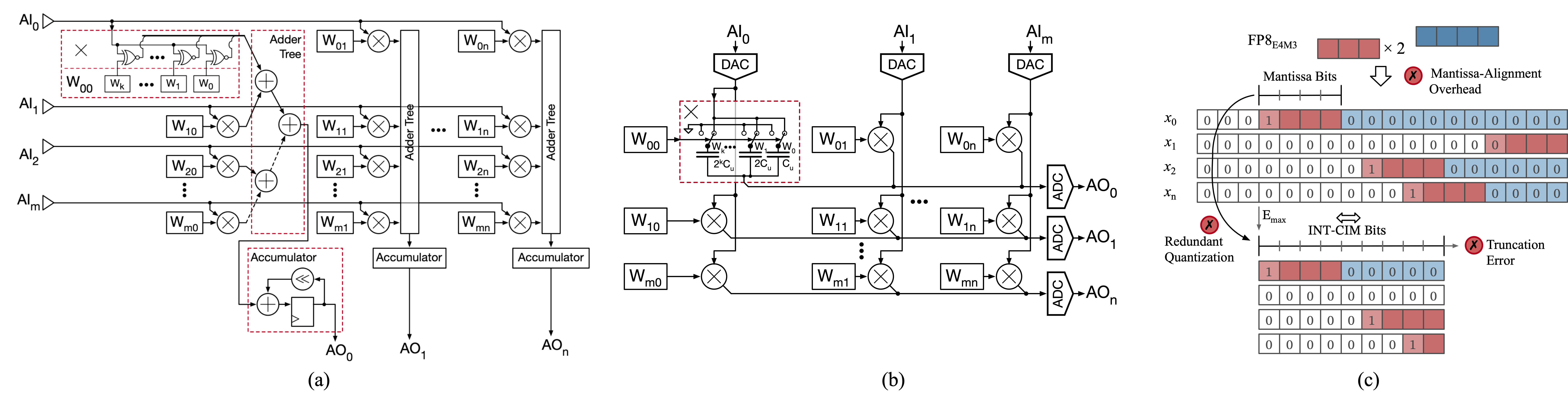}
	\caption{Conventional CIM based on (a) digital and (b) charge-based analog techniques. (c) Global normalization procedure used to align floating-point inputs for processing in conventional CIM arrays.}
	\label{fig:conv}
\end{figure*}

\subsection{Conventional Integer CIM}
Among many proposed analog and digital INT-CIM designs, the adder-tree-based digital CIM \cite{tsmccim22nm, flexdcim} and the capacitive-divider-based charge-domain analog CIM \cite{c2ccim, flexcim} stand out for their combination of energy-efficiency and robustness.

\subsubsection{Digital CIM}
The digital CIM shown in Fig.~\ref{fig:conv} (a) operates by accumulating partial products within an adder tree integrated into the SRAM array \cite{tsmccim22nm, flexdcim}. Distinct from standard memory access, wordlines are repurposed to function as single-bit activation inputs. These inputs trigger bit-wise multiplication with the stored weight data, generating partial products that enter the adder tree and are collapsed into a single dot product result. The array operates in a bit-serial fashion, cycling through activation bits to shift and accumulate the partial outputs. This architecture minimizes data movement by resolving partial products of activations and weights directly within the memory cells, and immediately collapsing outputs in $\log_2n$ stages into a single accumulated dot product. The digital CIM's energy efficiency is constrained by the dynamic switching of concurrent bitline readouts, the digital adder tree, and the accumulators. The digital INT-CIM's energy consumption scales quadratically with precision, as higher precision weights necessitate a proportional increase in accumulation logic, and higher precision activation inputs necessitate a proportional increase in bit-serial cycles, thus the compute energy mirrors the scaling of a conventional digital multiplier.

\subsubsection{Analog CIM}
To circumvent the digital CIM's high dynamic power cost, the analog CIM shown in Fig.~\ref{fig:conv} (b) replaces binary partial-product and accumulation logic with one-shot physical dot product computation in the charge domain \cite{c2ccim, flexcim}. By exercising a configurable network of capacitors, this architecture performs multiplication and summation directly in the charge domain. Capacitive dividers are interleaved into the SRAM array, where the locally stored weights configure the capacitor connections. DACs are the first step in the computation pipeline, driving voltages corresponding to a vector of activation inputs onto this capacitor network, allowing the dot product to be computed through charge redistribution, before being digitized by an ADC for storage and further processing. However, it is this digital-analog-digital conversion overhead that bounds the system's efficiency. Unlike the quadratic $N^2$ energy scaling of digital logic, the power consumption of a DAC and ADC scales exponentially with precision: following a $2^N$ trend in the technology-limited regime, and steepening to $4^N$ once thermal noise limits necessitate larger sampling capacitors\cite{adcsurvey}. Consequently, the viability of analog CIM is largely restricted to the low-precision domain, up to 8 bits \cite{murmann-mscim}, before exponential energy scaling erodes the efficiency benefits of one-shot parallelism.

\subsubsection{Limitations of Integer CIM}
Despite their architectural distinctions, both digital and analog INT-CIMs share a limitation: the rigidity of the integer data format. To accommodate a required dynamic range $DR$, an integer format necessitates a bit-width of $N \ge \log_2 DR$. As established above, energy penalties scale aggressively, making it expensive to simply widen the integer format to capture outlier-heavy distributions.  Furthermore, integer quantization enforces uniform resolution across the entire range, resulting in low information density for the various distributions typical of neural network weights and activations.

\subsection{Floating-Point CIM: Motivation, Approaches, and Challenges}
Motivated by the limitations of integer formats, recent CIM research has pivoted toward floating-point formats, which offer a choice of exponent and mantissa bits to allocate information to a logarithmic and linear scale, respectively. This extra flexibility allows designers to choose a format better suited to their application needs. However, realizing this advantage in CIM hardware presents an implementation challenge, as the exponent breaks the bit-alignment structure that benefits INT-CIM architectures. Whereas integer bits hold fixed significance, the significance of a floating-point mantissa bit is variable and depends on its exponent. This dependency prevents the direct in-situ accumulation of partial products common in integer arrays, forcing designers to adopt architectural workarounds that compromise the CIM's energy efficiency. Existing methodologies to address this challenge are discussed below.

\subsubsection{Distributed Digital Pipeline CIM}
One approach is to embed standard digital logic gates directly within the memory array, effectively distributing a digital multiplier pipeline across the SRAM rows \cite{digfpcim2}. This tactic is a floating-point analogy to the digital adder-tree INT-CIM presented earlier. However, because the exponent and mantissa must be processed together, the INT-CIM's bit-serial scheme with direct readout of compact 10T SRAM bitcells is replaced with heterogeneous SRAM bitcells and periphery that collectively implement a full FP multiplier for each 8-bit weight cell. As a result, SRAM area comprises only 2\% of the overall macro area and energy efficiency is $5.5\times$ worse than the same architecture's  \cite{digfpcim2} shared-exponent mode which functions similarly to an INT-CIM with a shared scaling factor.

\subsubsection{FP$\to$INT Conversion}
A popular strategy is to perform an FP-to-INT conversion to allow the use of a conventional INT-CIM for MAC operations \cite{\aligncimrefs, fpcim}. This is typically achieved through mantissa alignment \cite{\aligncimrefs}: First, the maximum exponent within a data block is identified, and individual values are denormalized by applying their exponent difference as a mantissa shift ($M_i \ll E_{\mathrm{max}}  - E_i$). This shift restores the bit-alignment property required for direct accumulation, thereby enabling INT-CIM techniques. However, this approach introduces a new inefficiency compared to conventional digital FP logic. A standard FP multiplier maintains efficiency by processing the mantissa and exponent separately, using a compact multiplier sized for the normalized mantissa and adders for exponents in the logarithmic domain. By contrast, after FP-to-INT conversion, the precision of the multiplier must be increased to fit the mantissa plus the range over which it may shift. This scales poorly with exponent bits due to the logarithm's compression: a digital multiplier's energy scales as $N^2 \rightarrow (2^{E_{\mathrm{bits}}})^2$, and analog data converters scale at up to $4^N \rightarrow 4^{2^{E_{\mathrm{bits}}}}$. Due to the severity of this penalty, lower-order bits are typically truncated, introducing a new energy-error trade-off. Furthermore, the FP-to-INT conversion results in high bit sparsity, as most bit positions are occupied by 0s representing the mantissa's position rather than its information. This is inefficient for analog processing, as DAC and ADC hardware costs are dictated by the global dynamic range rather than the active information content of the signal. To mitigate this, designers often adopt heterogeneous schemes to relax the precision requirements of the CIM array, such as redistributing sensitive MSBs to digital logic \cite{fphybrid} or offloading outliers to an auxiliary digital core \cite{intensive-cim-sparse-digital}. The compute overhead of mantissa alignment also introduces an energy bottleneck, for example, accounting for 39.8\% of overall BF16 MAC energy in \cite{fpaligncim}. This is typically minimized by pre-aligning weight mantissas offline and storing the maximum weight exponent $E_{\mathrm{max},W}$, as demonstrated in \cite{fpcim, fpaligncim}. However, runtime exponent normalization remains unavoidable for activation inputs. While an approach proposed by Y. Zhao et al. \cite{fpcim} introduces an FP-DAC to eliminate the input mantissa shift, the fundamental mismatch between overall FP dynamic range and INT-CIM precision scaling persists. Finally, some CIM works segment multi-bit accumulation compute lines into multiple compute lines binned by bit significance \cite{fpcim}; this lowers the peak ADC resolution requirement at the expense of additional conversions.

\subsubsection{Outlier-Aware}
S. He et al. propose an outlier-aware quantization and CIM \cite{oacim}, where most weights and inputs are quantized to INT4 while reserving FP16 capability for a small number of outliers, by provisioning an array of configurable MACs: either operating on a single 16-bit FP16 input, or four 4-bit INT4 inputs. This approach imposes a structural requirement on the input data, requiring that outliers account for no more than 3.125\% of the overall data. Therefore, the INT4 format must accommodate 96.875\% of the distribution. Additionally, configuring the MAC for an FP16 value requires pruning three adjacent INT4 values, although the outlier-aware quantization algorithm demonstrates robustness to this noise for the OPT-1.3B-to-13B family of models.

\subsubsection{Addition-Only}
Another approach, the addition-only CIM \cite{addcim}, approximates floating-point multiplication by removing the second-order term of the mantissa product: \((1.M_x \cdot 1.M_W) = (1 + M_x)(1 + M_W) = (1 + M_x + M_W + \cancel{M_x M_W})\). The computationally expensive $M_x \cdot M_W$ is removed, introducing a bounded error of at most $\nicefrac{1}{4}$, and smaller than $\nicefrac{1}{4}$ in most cases. A similar concept is used in \cite{lamfpcim}, replacing mantissa multiplication with addition for a logarithmic approximation of FP multiplication.

\vspace{1em}
In summary, while adopting floating-point formats in CIM architectures offers improved information efficiency, fully realizing this potential in hardware remains constrained by the structural mismatch between variable exponents and fixed memory arrays. Prior implementations have addressed this by making compromises on (1) energy of distributed digital pipelines \cite{digfpcim2}, (2) overheads and inefficiencies inherent to FP-to-INT conversion \cite{\aligncimrefs, fpcim}, (3) structural limitations on input data \cite{oacim}, and (4) fidelity loss associated with approximation schemes \cite{addcim, lamfpcim}.

\section{An Analog Gain-Ranging MAC and FP-CIM Architecture}\label{sec:arch}
We propose a \textit{Gain-Ranging MAC (GR-MAC)} to establish a new upper bound on the energy efficiency of the charge-based analog MAC and eliminate the structural mismatch between variable exponents and direct accumulation.

\begin{figure}[!t]
	\centering
	\includegraphics[width=3.5in]{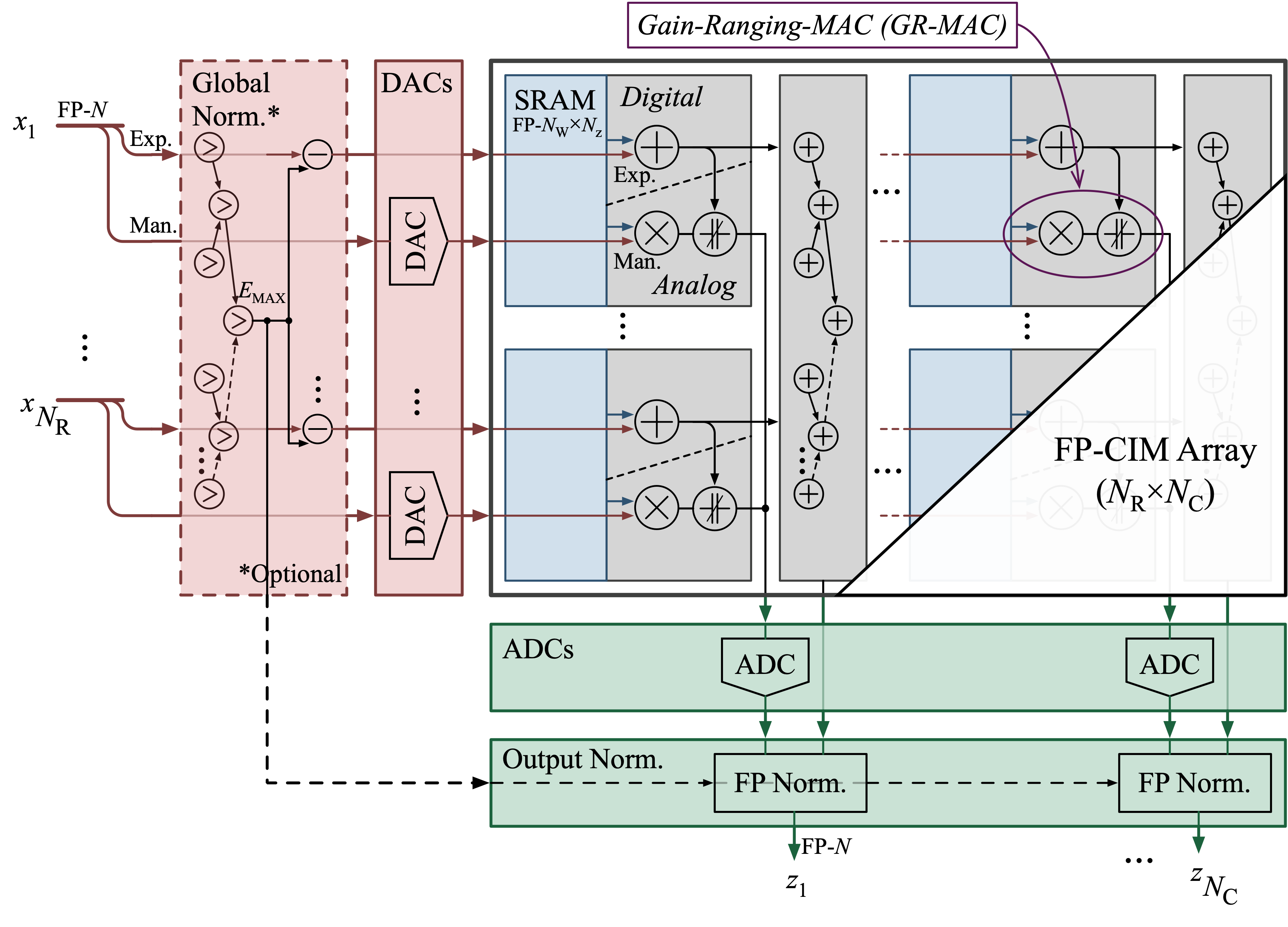}
	\caption{The proposed Gain-Ranging MAC and CIM architecture template for design space exploration. Global normalization \textit{(dashed lines)} is optional when more input dynamic range is required than the CIM array's native capability.}
	\label{fig:arch}
\end{figure}

Figure \ref{fig:arch} outlines the parameterized architecture for design space exploration. The mixed-signal CIM is organized as an array with $N_\mathrm{R}$ rows and $N_\mathrm{C}$ columns. Each row takes an $N$\nbhyphen bit floating-point input $x$ (FP-$N_x$, partitioned into $N_{\mathrm{E},x}$ exponent and $N_{\mathrm{M},x}$ mantissa bits), typically representing a neural network activation. The input's exponent is broadcast digitally, and the mantissa is distributed as an analog voltage via a DAC. Within the array, each unit cell packs $N_{\mathrm{Z}}$ FP\nbhyphen $N_W$ weight values in local SRAM, along with switched-capacitor structures for charge-based MAC and an $N_E$\nbhyphen bit adder to compute the output exponent for gain ranging and normalization. The cell contributes its analog MAC output charge to its column's shared accumulation line, and a digital adder tree in each column accumulates the cell's output exponents encoded in a one-hot magnitude format $(1 \ll E_i)$ to determine the column's overall normalization factor. At the column periphery, the analog output is quantized by an ADC, which is then multiplied and normalized with the column normalization factor to obtain the final dot product result. Optionally, the CIM array is wrapped in a global exponent normalization block to enable a wider input dynamic range than the CIM's native capability. Global normalization operates on the same principle as conventional FP-to-INT transfer CIMs \cite{\aligncimrefs, fpcim}, but with an energy and fidelity overhead due to the additional compute and truncation.

\subsection{Notation}
We define the following conventions to facilitate our discussion and analysis. Boldface characters denote vectors and matrices (e.g., inputs $\bm{x}$, weights $\bm{W}$, pre-accumulation products $\bm{p}$, and pre-activation outputs $\bm{z}$), while standard italics denote the scalar elements processed by individual unit cells and data converters (e.g., $x, W, p, z$). Signal variables are treated as dimensionless quantities normalized to the unit interval $[-1,+1]$ for a signed signal and $[0,1]$ for an unsigned signal; physical values are easily obtained by scaling to a full-scale reference (e.g., $V_{x} = x V_{\mathrm{FS}}$). To resolve ambiguity between overlapping terms, we distinguish the floating-point exponent, $E$, from the energy consumption metric, $\mathcal{E}$. The stored components of a floating-point value are denoted as $S$ (Sign), $E$ (Exponent), and $M$ (Mantissa). However, in the context of physical definitions, $M$ and $E$ denote the effective values rather than raw bit fields. $M$ represents the significand normalized to the unit interval, $M_x = 1.M_{x, \mathrm{stored}}/2 \in [0.5, 1)$ for normals, and $M_x = 0.M_{x, \mathrm{stored}}/2 \in [0, 0.5)$ for subnormals). Likewise, the reserved code for subnormals is resolved for $E$, $E_x = \max(1, E_{x, \mathrm{stored}})$. Thus, the value of a floating-point scalar $x \in [-1,+1]$ is easily written: $x = (-1)^{S_x} \cdot M_x \cdot 2^{E_x - E_{\mathrm{max}}}$.

\subsection{The Gain-Ranging MAC (GR-MAC)}
The charge-domain CIM array is a network of configurable capacitors that performs a Matrix-Vector Multiplication, $\bm{z} = \bm{Wx}$. The array comprises unit cells, each performing a single multiply-accumulate (MAC) operation. Each weight, $W$, is stored in a unit cell and used to configure the local capacitive divider, thus embedding $\bm{W}$ in the array. The input, $\bm{x}$, is applied across the rows of the array by DACs as voltages. Charge redistribution through the capacitive network generates the output voltages, $\bm{z}$, which are then quantized by ADCs at the column periphery. The ADC energy consumption scales with the required precision, $\mathrm{ENOB}$, and becomes the primary energy bottleneck as the charge-based CIM scales \cite{murmann-mscim}. The ADC energy scaling is modelled as $\mathcal{E}_{\mathrm{ADC}} \propto \mathrm{ENOB} + \gamma 4^\mathrm{ENOB}$, representing a linear baseline energy per conversion step, and a $4^N$ term for thermal-noise-limited scaling, consistent with the SAR topology commonly used in CIM designs \cite{sun-aord}. The crossover point between the two scaling regimes is defined by $\gamma \approx \nicefrac{N_{\mathrm{cross}}}{4^{N_{\mathrm{cross}}}}$ and $N_{\mathrm{cross}}\approx  10$~bits in \cite{murmann-mscim}, acting as a practical boundary for energy consumption.

The following subsection, \ref{sec:shrinkydinky}, explains how dot product computation on data that is globally normalized to a block-wise maximum, such as an aligned integer format \cite{\aligncimrefs} or FP-DAC \cite{fpcim}, produces a \textit{signal shrinkage} \cite{murmann-mscim} effect. This forces the ADC to satisfy an \textit{excess resolution requirement} \cite{murmann-mscim} to lower the noise floor to distinguish the compressed signal, at an energy cost up to $4^N$. By contrast, the GR-MAC leverages the separation of range and resolution in floating-point formats to apply \textit{per-unit normalization}. Subsection \ref{sec:loudandproud} explains the operation of the GR-MAC, showing how weighted accumulation via gain-ranging replaces uniform averaging to minimize signal shrinkage, thereby relaxing the ADC requirements.

\subsubsection{INT-MAC and Signal Shrinkage}\label{sec:shrinkydinky}
\begin{figure*}[!t]
	\centering
	\includegraphics[width=7.16in]{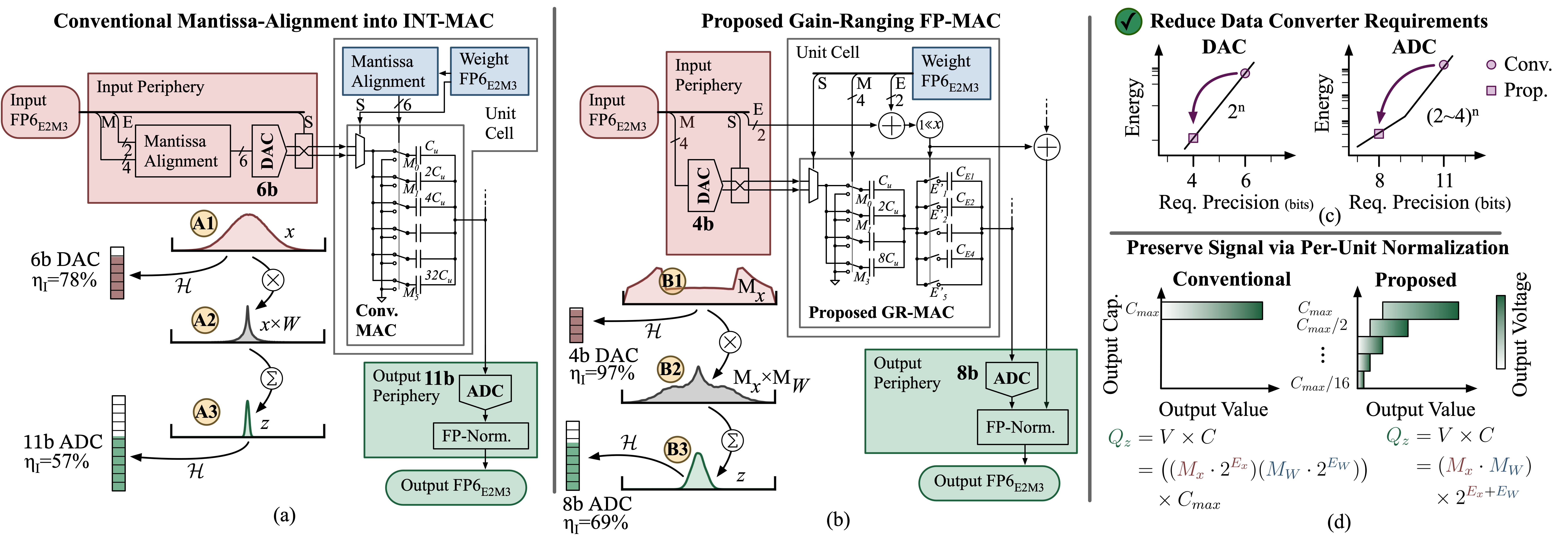}
	\caption{(a) Conventional and (b) proposed MAC unit with periphery, enabling (c) a reduction in data converter precision, and (d) an illustration of the magnitude-based binning of normalization. The effect of redundant precision in the DAC and ADC required by conventional FP-to-INT architectures is shown as bars filled to represent the information content of the quantized inputs as a fraction of the capacity dictated by DAC and ADC requirements. A normal distribution clipped to $4\sigma$ is chosen for the input $x$ and weight $W$ to generate the distributions A1..B3 using a Monte Carlo simulation; the proposed GR-MAC's benefit is not strictly reliant on this input distribution. Similarly, $N_{\mathrm{R}}=32$ and the FP6 data format are chosen as an example for illustration. The ADC resolutions are specified according to the analysis in Section \ref{sec:anal}.}
	\label{fig:comp}
\end{figure*}

The conventional multi-bit charge-based INT-MAC, shown in Fig.~\ref{fig:comp}~(a), works by scaling the input voltage through a capacitive divider with a ratio configured by the weight, performing a local $p_{i,j} = x_i \cdot W_{i,j}$, which is connected to a shared column compute line for accumulation ($z_j = \nicefrac{1}{N_{\mathrm{R}}} \sum_{i=1}^{N_{\mathrm{R}}} p_{i,j}$). The INT-MAC's output signal power is compromised by signal shrinkage due to the statistics of multiplication and accumulation within a fixed full-scale range. Since the output scale is fixed to accommodate the maximum product ($1 \times 1$), the multiplication of fractional inputs ($|x|,|w| \le 1$) contracts the signal variance relative to the full-scale limit. Adopting a first-order model of uncorrelated, zero-mean inputs, this contraction is given by $\sigma^2_{p} = \sigma^2_x \sigma^2_W$, as illustrated in Fig. \ref{fig:comp} \circled{A1} $\rightarrow$ \circled{A2}. Signal variance is further suppressed by accumulation; each MAC's output remains connected to the column compute line, physically realizing the accumulation as dividing each MAC's charge across the total lumped capacitance. While this inherent averaging accommodates the theoretical worst-case sum without saturation, variance further shrinks by the column depth, $\sigma^2_{z} = \sigma^2_{z,i}/N_\mathrm{R}$, as shown in Fig. \ref{fig:comp} \circled{A2} $\rightarrow$ \circled{A3}.

Crucially for FP-to-INT architectures, this signal degradation is compounded by the resolution mismatch created by mantissa alignment during global normalization. By forcing local inputs to align with a block-wise global maximum, the required integer bit-width is set to accommodate a range of exponents, thereby demanding more accuracy than is strictly specified by the floating-point mantissa.

The ADC minimum energy bound is ideally set by the requirement to keep noise introduced by analog processing below the quantization noise floor inherent to the input's format. However, signal shrinkage introduced by the INT-MAC operation reduces signal power, requiring a commensurate suppression of the noise floor to maintain the target SNR during output quantization. This is quantified as an excess resolution requirement, raising the ADC energy per conversion by a factor of up to $4^{\mathrm{ENOB}}$. Additionally, the increased bit-width caused by FP-to-INT conversion increases the required precision of the input DACs and further attenuates the output signal according to the block-wise maximum of the input data. The next section explains the operation of the GR-MAC and showcases how per-unit normalization relaxes the ADC requirement.

\subsubsection{GR-MAC and Signal Preservation}\label{sec:loudandproud}
Floating-point formats offer a natural foundation to address this bottleneck, as range (DR) is no longer strictly tied to precision (SQNR), with $N_E$ and $N_M$ allowing for decoupled control of each. This decoupling eliminates the INT-MAC's requirement to pay for a high DR with large integer bit-widths, which previously inflated analog resources. Instead, the proposed GR-MAC leverages this inherent separation of range and resolution to apply \textit{per-unit normalization}. This eliminates the need to compress the entire global dynamic range into the fixed voltage scale and offloads the low-overhead exponent arithmetic to the digital domain.

The GR-MAC operation is shown in Fig.~\ref{fig:comp} (b). The multiplier functions similarly to the INT-MAC, with a digital weight input configuring a capacitive divider to operate on an analog voltage activation input. However, the GR-MAC multiplier operates only on the normalized floating-point mantissas, \circled{B1}, instead of denormalized integer values, \circled{A1}. This produces a normalized output product, \circled{B2}, which cannot be directly distributed onto the column compute line because its exponent may differ from that of other outputs. This is the conflict that prevents direct-accumulation and why other architectures must denormalize inputs before multiplication. Instead, the GR-MAC introduces a gain-ranging stage to enable \textit{weighted accumulation}: Outputs are shared, scaled according to their exponent, by selecting a variable capacitive coupling to the column compute line as $C_{\mathrm{cpl}} \in \left\{C_{\mathrm{tot}}, \frac{C_{\mathrm{tot}}}{2}, \frac{C_{\mathrm{tot}}}{4}, \dots, \frac{C_{\mathrm{tot}}}{2^{E_{\mathrm{max}}-1}}\right\}$, where $C_{\mathrm{tot}} = (2^{N_{\mathrm{M},W}}-1) \cu$ represents the total capacitance of the mantissa multiplication stage and therefore the maximum available coupling capacitance. The switched-capacitive coupling stage is digitally controlled, using a one-hot input computed as the sum of the input and weight exponents, $1 \ll E_x + E_W$. Coupling capacitors are sized to create the desired ratios, such that $C_{\mathrm{E},i} \parallel C_{\mathrm{tot}} \propto 2^{E_i}$. While the conventional MAC's column compute line output is available purely in the voltage domain, as its charge is always scaled to a constant worst-case $N_\mathrm{R} C_{\mathrm{tot}}$, the GR-MAC column output capacitance varies, keeping the output voltage normalized for efficient ADC conversion. A digital adder tree sums the one-hot output exponents to obtain the total column capacitance: this is the overall scaling factor to which the column output is normalized after ADC conversion. The adder tree cycles once per MVM and operates on low-activity one-hot inputs. By contrast, a conventional bit-serial digital CIM architecture (Fig.~\ref{fig:conv} (a)) requires one cycle for each input bit position.

The following subsection explains how the GR-MAC preserves the output signal to reduce the ADC's $\mathrm{ENOB}$ requirement, thereby improving energy efficiency. First, while the effect of multiplication ($\sigma^2_{a\cdot b} = \sigma^2_a \sigma^2_b$) is unavoidable, floating-point normalization maximizes the input amplitude by definition. This effect is seen when comparing the conventional integer input [A1] to the proposed mantissa input \circled{B1}, leading to a wider product distribution in \circled{B2} than in \circled{A2}. While normalized inputs maximize the initial signal swing, the dominant source of shrinkage in the conventional INT-MAC is the accumulation stage, where uniform averaging suppresses variance by the column depth $\sigma^2_{z} = \sigma^2_{z,i}/N_\mathrm{R}$. The GR-MAC directly addresses this by replacing uniform averaging with exponent-weighted averaging. This advantage is quantified by replacing $N_\mathrm{R}$ with an effective number of contributors, $N_{\mathrm{eff}}$, to govern shrinkage. Adopting the standard formulation for weighted samples, this is defined as $N_{\mathrm{eff}} = \nicefrac{\left( \sum 2^{E_i} \right)^2}{\sum 4^{E_i}} \le N_{\mathrm{R}}$. While the worst-case scenario ($N_{\mathrm{eff}} = N_\mathrm{R}$) occurs only when all inputs and weights share the same exponent value, practical data distributions exhibit significant variation. In these cases, the exponential weighting allows the largest signals to dominate the average, driving $N_{\mathrm{eff}} \ll N_\mathrm{R}$. The example conditions in Fig.~\ref{fig:comp} produce $N_{\mathrm{R}}=32$ in \circled{A3} and $N_{\mathrm{eff}} = 14.6$ in \circled{B3}, and combined with normalized inputs improves the overall output signal power by $20 \times$, for a reduction in the ADC excess resolution requirement of $\Delta \mathrm{ENOB} = 2.2$ bits. This benefit is quantified across different formats and input conditions in Section \ref{sec:anal}.

Overall, the benefit of the GR-MAC is summarized as overcoming the rigid coupling between dynamic range and efficiency inherent to conventional analog INT-MAC. By processing normalized signals, the architecture maximizes the utilization of available signal swing, thereby minimizing the excess resolution requirement that drives ADC energy.

\subsection{Architecture Flexibility and Parameterization}
\begin{figure}
	\centering
	\includegraphics[width=1.5in]{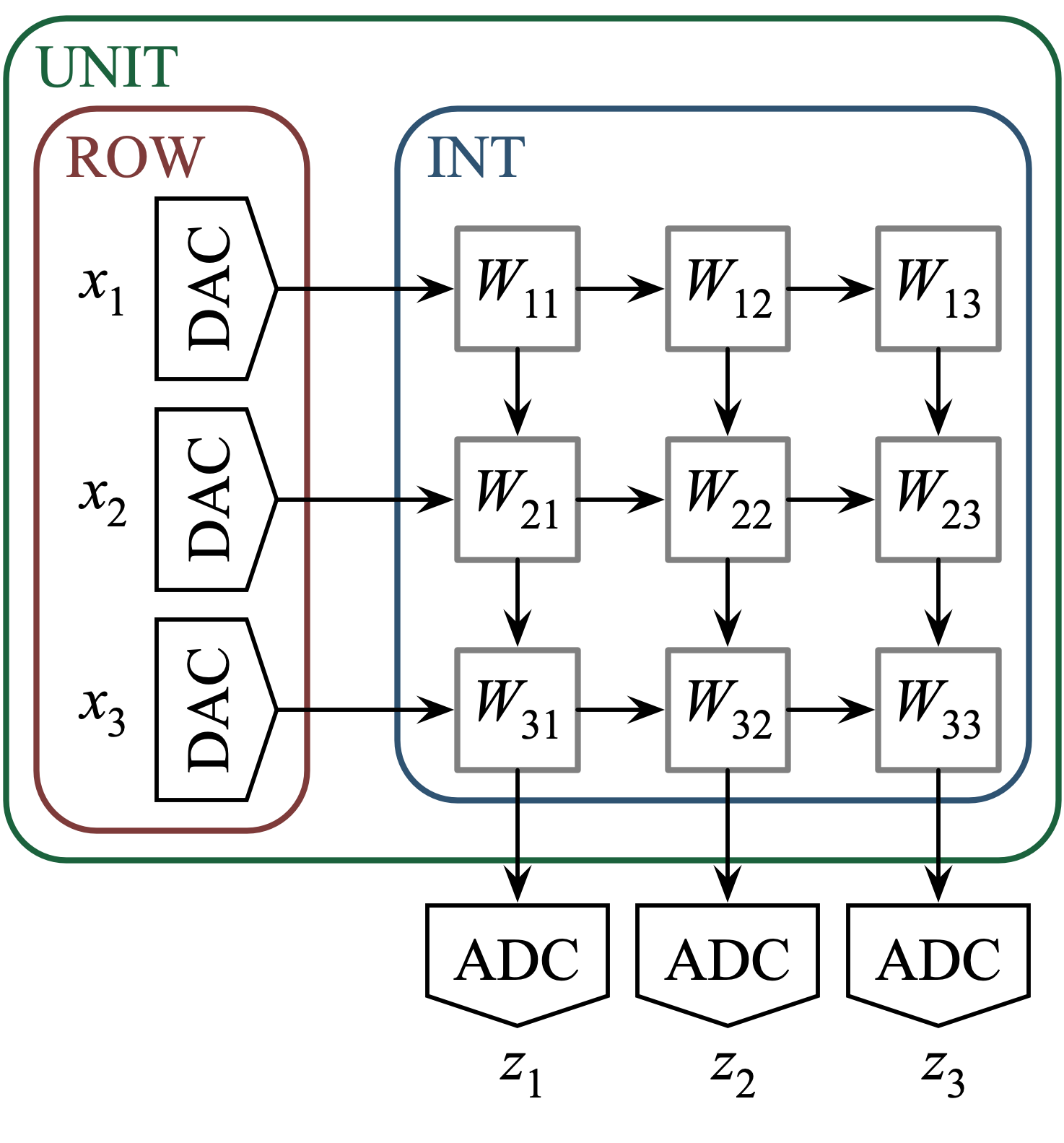}
	\caption{Possible normalization domains for a GR-MAC-based architecture. The input or weight values within each boundary are normalized and carry exponents for that normalization granularity.}
	\label{fig:normtypes}
\end{figure}

Different granularities of normalization may be chosen, varying the trade-off between maximizing signal amplitude and complexity of exponent processing. This section describes these options and the digital components which handle exponent bookkeeping. The domains for different normalization granularities are illustrated in Fig.~\ref{fig:normtypes}.

\subsubsection{Unit Normalization} Up to this point, the GR-MAC is described according to per-unit normalization, where each cell's output capacitance is scaled by both the input and weight exponents, thereby providing the most fine-grained normalization. Since each cell has a unique output exponent, an adder tree is required for each column. Each unit cell also contains a digital adder and binary decoder to generate the output exponent and switch configuration for the gain-ranging stage. Finally, the dot product output is produced by multiplying the column-compute-line voltage converted by the ADC with the column exponent sum produced by the adder tree. The adder trees and output multiplication and normalization are amortized over each column by $N\mathrm{_R}$, and the unit cell logic is not amortized. Unit normalization is most energy efficient when the baseline ADC resolution requirement is high. In Section \ref{sec:anal}, the efficiency crossover point is identified at $N_{\mathrm{M},x} \ge 6$ in 28 nm technology; scaling to advanced nodes would likely lower this threshold due to superior digital scaling. A further limitation of unit normalization is that the bounded dynamic range of the gain-ranging stage must accommodate the sum of the input and weight exponents.

\subsubsection{Row Normalization} The normalization bookkeeping overhead can be significantly reduced by only applying normalization to the inputs at run-time. Then, since the output capacitance of each unit cell depends only on the input exponent, the unit cell is greatly simplified: The local exponent adder is removed, and the exponent decoder serves an entire row instead of a single unit. Furthermore, only one exponent adder tree is required for the entire CIM array, and its input bitwidth is reduced. The exponent decoder is amortized over each row by $N\mathrm{_C}$, the exponent adder tree is amortized over the entire CIM array by $N\mathrm{_R} \times N\mathrm{_C}$, and column output multiplication and normalization are unchanged, amortized over each column, by $N\mathrm{_R}$. The weights may be either stored shifted at the expense of storage overhead \cite{fpcim}, or shifted at run-time at the expense of logic overhead for exponent decoding and mantissa shift. Section \ref{sec:anal} shows that row normalization granularity is most energy efficient when the ADC operates well below the thermal-noise-limited scaling regime in 28 nm technology. Finally, as the entire dynamic range of the gain-ranging stage is allocated towards the input, row-normalization offers a higher input dynamic range ceiling than unit normalization.

\subsubsection{INT Normalization} Integer inputs lack an exponent to provide scale information; however, if floating-point weights are used, a normalization benefit may still be obtained. Column exponent sums are pre-computed at compile-time, reducing digital logic to the bare minimum: an exponent decoder in each unit cell drives the switches in the gain-ranging stage, and a multiplier at each column output produces the final result.

\subsection{GR-MAC Implementation Considerations}
Although the proposed GR-MAC architecture, shown in Fig.~\ref{fig:arch}, focuses on establishing the theoretical limits of analog gain-ranging computation, physical realization of the circuit faces practical challenges similar to those in classical C-2C and binary-weighted DAC ladders. In particular, introducing a switched capacitive coupling stage to the output of a binary-weighted capacitor array leads to two key implementation concerns: (1) the need for non-multiple capacitor sizing to achieve the desired exponential gain profile, and (2) the introduction of parasitics due to floating top and bottom plates. Fortunately, both issues are well-studied in the context of C-2C DACs and compact capacitor array implementations, and several established techniques apply to GR-MAC design \cite{c2ccim, c2c-razavi, cap-finger-layout}. The main considerations are summarized below.

\begin{figure}
	\includegraphics[width=3.2in]{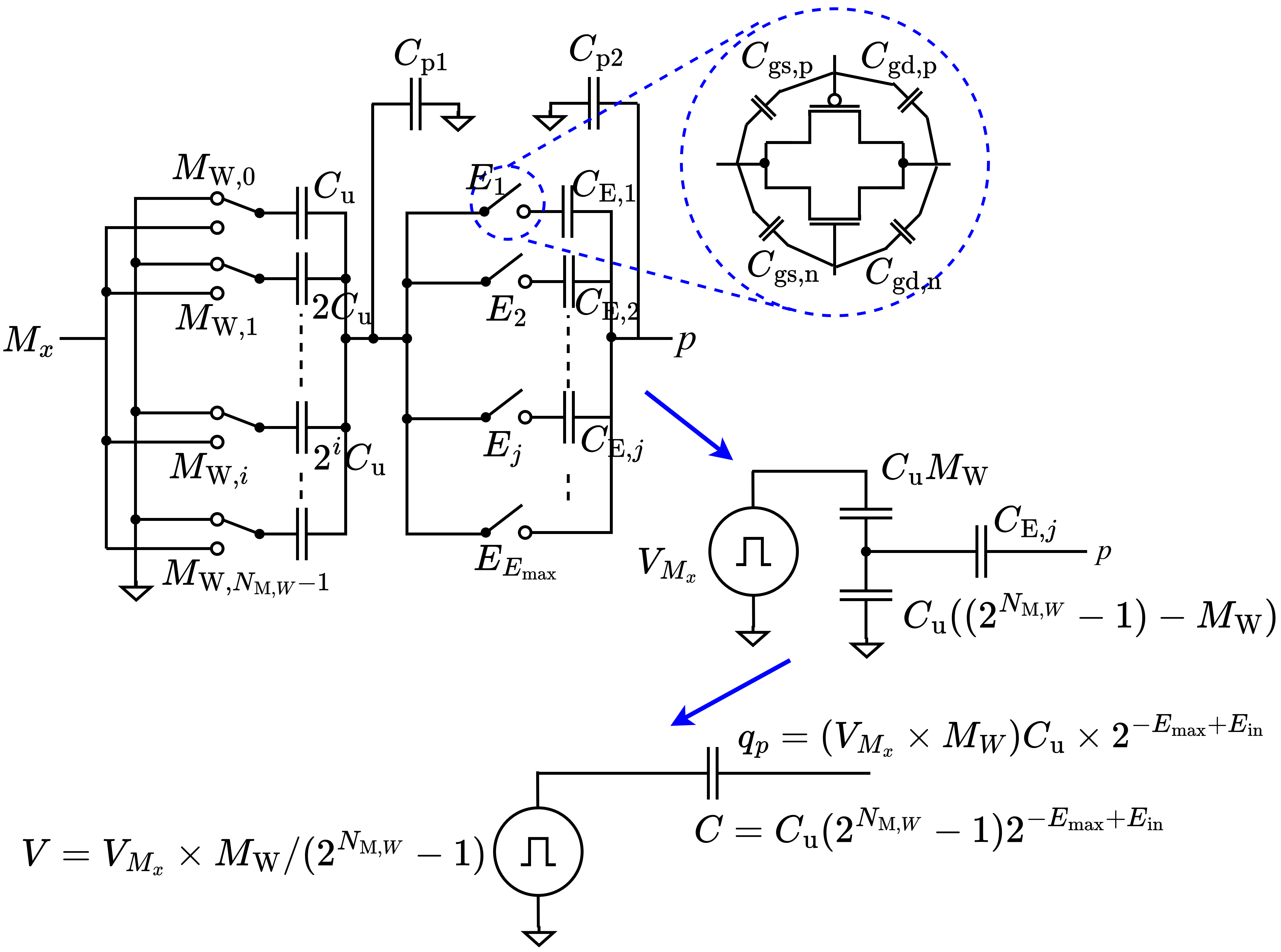}
	\caption{Switched-capacitor divider with variable output capacitance used for GR-MAC. Switch and floating top- and bottom-plate parasitics are modelled to a first-order approximation as lumped capacitances $C_{\mathrm{p}1}$ and $C_{\mathrm{p}2}$ at the floating nodes. The GR-MAC is simplified into an equivalent circuit of a single capacitively-coupled voltage source. The voltage source becomes a function of the multiplication of the input mantissas, and the capacitive coupling is written as a function of the exponent raised to the power 2.}
	\label{fig:implem}
\end{figure}

\subsubsection{Parasitic Capacitance Compensation} As illustrated in Fig. \ref{fig:implem}, parasitic capacitances perturb the intended division ratio of the coupling network, degrading linearity and altering the gain boundaries between ranges. A well-known remedy in C-2C DACs \cite{c2c-razavi} is to slightly enlarge the branch capacitors so that the effective division ratio, including parasitics, returns to an ideal $\nicefrac{1}{2}$ at each stage. In this approach, the branch capacitors are increased to $\alpha C = C_\mathrm{p}$, absorbing the parasitic into the capacitor sizing, restoring the recursive halving condition. We similarly derive a closed-form expression for the required resizing of coupling capacitors in the GR-MAC, effectively cancelling the effect of $C_{\mathrm{p}1}$, while $C_{\mathrm{p}2}$ is absorbed into the overall compute line capacitance and does not affect linearity:
\begin{equation}
C_{\mathrm{E}_j} = \frac{\left(2^{N_{\mathrm{M},W}+1}-1\right)C_\mathrm{u} + C_{\mathrm{p1}}}{2^{E_{\max}-E_j} - 1} \label{eq:capsize}
\end{equation}
Because the correction depends directly on extracted parasitics, its effectiveness relies on accurate post-layout extraction. Prior work in INT-CIM macros demonstrates that this technique is sufficient to maintain linearity in 8-bit C-2C ladders fabricated in advanced FinFET processes \cite{c2ccim}. The same methodology applies naturally to the GR-MAC’s gain-ranging capacitive network, and a conservative limit of 6 bits is assumed for the analysis in Section \ref{sec:anal}.

\subsubsection{Arbitrary Capacitor Sizing}
The gain-ranging stage requires capacitor values determined by eq. \ref{eq:capsize}, which generally cannot be realized using integer multiples of a unit capacitor. To accurately implement these non-integer ratios without introducing mismatch from fractional-unit structures, the unit-length scaling layout technique from \cite{cap-finger-layout} can be adopted. Since the capacitance scales linearly with metal length, this method enables fine-grained, monotonic tuning with resolution limited only by lithography, while preserving excellent matching and compact area.

\subsection{GR-MAC Post-Layout Simulation and Mismatch Sensitivity}
To validate the GR-MAC, the FP6\textsubscript{E2M3} configuration introduced in Fig.~{\ref{fig:comp}~(b) is implemented in a 22 nm planar process, as shown in Fig.~\ref{fig:layout}. To aid physical implementation, two circuit transformations are applied. First, the minimum coupling switch is removed so that $C_\mathrm{E1}$ always couples to the output; the value of $C_\mathrm{E1}$ is subtracted from $C_\mathrm{E2..4}$. Next, the largest exponent, $E_\mathrm{4}$, activates both $C_\mathrm{E3}$ and $C_\mathrm{E4}$, shrinking the size of the largest capacitor, $C_\mathrm{E4}$. These optimizations reduce the overall implementation area without sacrificing any degrees of freedom for tuning. The capacitor values used for this implementation are provided in Table~\ref{tab:caps}, with the ideal schematic values, extracted capacitor values after initial layout, and final extracted values after adjusting the finger lengths of $C_\mathrm{E1..4}$. The largest change is seen in the decrease of $C_\mathrm{E1}$, by absorbing the mutual coupling between the large first-stage divider output net, and second-stage coupling output net.

\begin{figure}
		\centering
		\includegraphics[width=2.6in]{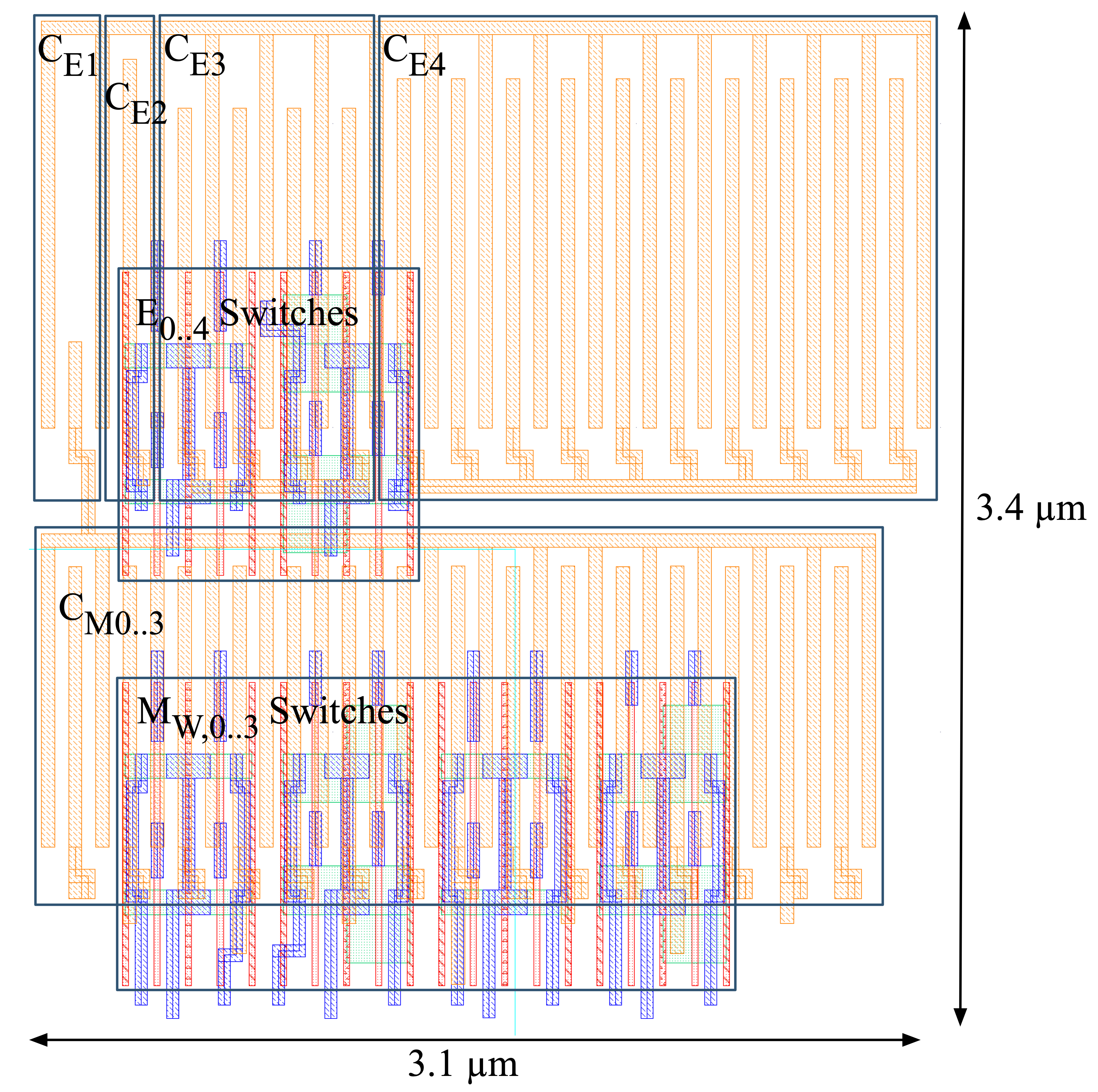}
		\caption{Layout of the GR-MAC capacitive structure and switches in the FP6\textsubscript{E2M3} configuration. A 22 nm planar process is used with 5-layer interdigitated metal-oxide-metal (MOM) capacitors for both the divider (C\textsubscript{M0}..C\textsubscript{M3}) and variable coupling stage (C\textsubscript{E0}..C\textsubscript{E3}). Only the first layer is shown for clarity; upper layers are jogged laterally by a half-unit-capacitor-width to maximize vertical coupling.}
		\label{fig:layout}
\end{figure}

\begin{table}[ht]
	\centering
	\caption{FP6\textsubscript{E2M3} GR-MAC Capacitor Values in 22 nm FD-SOI}
	\label{tab:caps}
	\begin{tabular}{@{}lccc@{}}
		\toprule
		\textbf{Capacitor} & \textbf{Schematic} & \textbf{Initial} & \textbf{Tuned}  \\
		                   & \textbf{(fF)}      & \textbf{Post-Layout (fF)}    & \textbf{Post-Layout (fF)}   \\
		\midrule
		$C_{\mathrm{M0}}$ & 1    & 0.94 & --             \\
		$C_{\mathrm{M1}}$ & 2    & 1.85 & --             \\
		$C_{\mathrm{M2}}$ & 4    & 3.72 & --             \\
		$C_{\mathrm{M3}}$ & 8    & 7.46 & --             \\
		$C_{\mathrm{E1}}$ & 1    & 1.03 & 0.42 ($-59\%$) \\
		$C_{\mathrm{E2}}$ & 1.14 & 1.06 & 1.23 ($+16\%$) \\
		$C_{\mathrm{E3}}$ & 4    & 3.71 & 4.19 ($+13\%$) \\
		$C_{\mathrm{E4}}$ & 10   & 9.32 & 11.4 ($+22\%$) \\
		\bottomrule
	\end{tabular}
\end{table}

\begin{figure}
	\centering
	\includegraphics[width=3.5in]{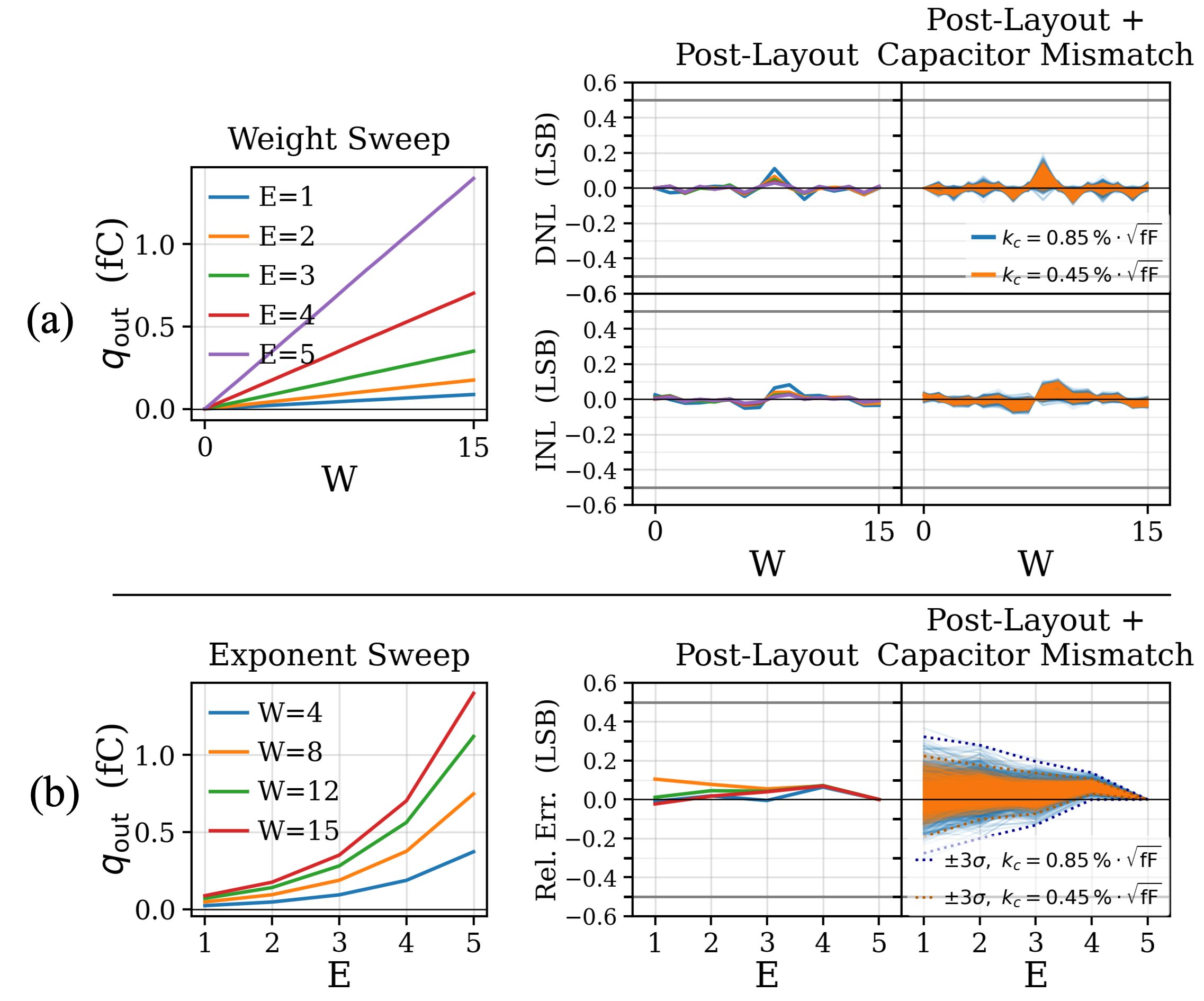}
	\caption{Post-layout simulation of the FP6\textsubscript{E2M3} GR-MAC shown in Fig.~\ref{fig:layout}. (a) W sweep at each E producing a linear response, and differential- and integer-non-linearity (DNL \& INL) under nominal conditions. DNL \& INL are also simulated under capacitor mismatch using mismatch factors derived in Sec.~\ref{sec:mismatch}, with $n=1000$. (b) E sweep across the range of W producing an exponential response. Relative error is plotted, normalized to the LSB step size of the W input for both the nominal case and under mismatch.}
	\label{fig:capsim}
\end{figure}

\subsubsection{Capacitor Mismatch}\label{sec:mismatch}

The GR-MAC's robustness to capacitor mismatch is evaluated using Monte Carlo analysis. Since the foundry does not provide mismatch models for fF-scale metal-oxide-metal (MOM) capacitors, the mismatch of each unit capacitor is modelled using Pelgrom's area relation~\cite{pelgrom}, which has been experimentally verified for MOM capacitors down to 0.45~fF in~\cite{cap_tripathi, cap_omran}. The mismatch standard deviation is given by $\sigma(\nicefrac{\Delta C}{C}) = \nicefrac{K_A}{\sqrt{A_{\mathrm{cap}}}}$, where $A_{\mathrm{cap}}$ is the total dielectric plate area and $K_A$ is a matching coefficient. Omran~et~al.\ measured $K_A = 0.48\%{\cdot}\mu\mathrm{m}$ in a $0.18~\mathrm{\mu m}$ process and confirmed its invariance to finger spacing~\cite{cap_omran}. Since both capacitance and plate area scale linearly with finger length for a given cross-section, this
can be rewritten as $\sigma(\nicefrac{\Delta C}{C}) = \nicefrac{K_C}{\sqrt{C}}$, where
$K_C = \nicefrac{K_A}{\sqrt{A_{\mathrm{eff}}/C}}$. The ratio $\nicefrac{A_{\mathrm{eff}}}{C}$ is a geometric factor determined by the capacitor cross-section, accounting for the different widths and spacings of the lateral and vertical dielectric regions that contribute to the overall capacitance. For the five-layer interdigitated capacitors used in this work, laid out at minimum spacing rules in 22~nm, $\nicefrac{A_{\mathrm{eff}}}{C} = 1.12~\mu\mathrm{m}^2/\mathrm{fF}$, yielding $K_C = 0.45\%{\cdot}\sqrt{\mathrm{fF}}$ (using the $K_A$ experimentally measured in \cite{cap_omran}). Tripathi~et~al.\ directly measured $K_C = 0.85\%{\cdot}\sqrt{\mathrm{fF}}$ for single-layer lateral finger capacitors in a 32~nm SOI process~\cite{cap_tripathi}. We estimate $K_C$ for our structure to lie in this range, $0.45$--$0.85\%{\cdot}\sqrt{\mathrm{fF}}$, and simulate both bounds in the Monte Carlo analysis shown in Fig.~\ref{fig:capsim}. The post-layout simulation under $3\sigma$ mismatch remains within the $\nicefrac{1}{2}$ LSB bound across both inputs. The highest mismatch sensitivity occurs at low E values due to the small output LSB step size; in a practical scenario with many contributions to the total column charge, the cumulative error margin is dominated by the large E values, thereby reducing this impact.

\subsubsection{GR-MAC Area Overhead}
The sample FP6\textsubscript{E2M3} GR-MAC layout in Fig.~\ref{fig:layout} has a unit-cell area of $10.5~\mathrm{\mu m}^2$. The addition of a gain-ranging stage to the base 4-bit capacitive divider increases the cell's output dynamic range by $2^4$ (4 bits), switch area by 50\% to $1.5\times$, and capacitor area by $2.6\times$. Scaling DR to the same amount by conventional means would require a $16\times$ increase in capacitor area. Under a row-normalization scheme, additional cell logic is not required. However, the unit-normalization scheme presented in Fig.~\ref{fig:comp} requires a 2-bit adder and 3-bit decoder for FP6\textsubscript{E2M3}; this logic consumes an overhead of $8.6~\mathrm{\mu m}^2$, and can be stacked underneath the capacitors as it does not toggle during computation.

\section{Analysis and Discussion}\label{sec:anal}
This section presents a design-space exploration to quantify the energy benefits which are obtained by adding gain-ranging to the analog MAC. We first analyze the ADC resolution, as the thermal-noise-limited regime boundary is the dominant scaling factor for CIM energy. Next, we use energy modelling to quantify the overall energy advantage of gain-ranging, including the overheads introduced by the additional digital logic required for exponent bookkeeping in the GR-MAC.

\subsection{ADC Requirements}

To ensure robust analog processing of a given floating-point input, we restrict the permissible noise introduced by the ADC to 6 dB under the quantization noise floor of the input format, aligning with the findings in \cite{murmann-mscim}. The relative error, and therefore SQNR, of a floating-point format is related to the number of mantissa bits $\mathrm{SQNR} \approx 6.02 N_\mathrm{M} + 10.79~\mathrm{dB}$ \cite{quantnoisebook}, and is independent of the input data distribution, provided the data remains within the format's range. Although the format offers a bounded SQNR target, its physical realization requires that the ADC noise floor decrease linearly with signal amplitude, which is data-dependent. Thus, for a rigorous ADC specification, it is necessary to consider both the SQNR requirement, and the input data distribution, with particular consideration of long-tailed distributions that can cause small output amplitudes. We evaluate three distinct distributions, illustrated in Fig. \ref{fig:sqnr}, to define the hardware requirements:

\begin{enumerate}[label=\roman*)]
	\item{\textit{Uniform Distribution (baseline)}} The uniform distribution is the standard baseline for conventional integer CIM analysis \cite{murmann-mscim}. However, later in this subsection, it is shown to underestimate the conventional CIM's ADC requirement for practical inputs.
	\item{\textit{Maximum-Entropy Distribution}} We consider the maximum-entropy distribution as the floating-point equivalent of the uniform INT baseline. Defined as the distribution matching the quantizer prior, it can be obtained by uniformly randomizing the bits of a given format. This distribution is useful because it represents optimality in the choice of data format for a given workload; this is the explicit objective in some quantization works \cite{qlora}. 
	\item{\textit{Gaussian + Outliers Distribution (Empirical Dynamic-Range Stress Test)}} To model the activation patterns empirically observed in large language models \cite{llmint8, smoothquant, awq}, we construct this distribution as a mixture of a Gaussian core with uniformly distributed high-magnitude outliers. This distribution is defined by the outlier probability, $\epsilon$, and magnitude, $k$, defined relative to the core's $3\sigma$. It serves as a stress test for dynamic range: the hardware must resolve the narrow core without underflow, while simultaneously accommodating rare, massive outliers without clipping. For this analysis, we select values of $\epsilon=0.01$ and $k=50$, consistent with empirical observations in \cite{llmint8, smoothquant, awq} regarding the sparsity and magnitude of emergent features.
\end{enumerate}

Next, the impact of the established distributions on quantization noise and ADC requirements for analog CIM computation is analyzed.

\subsubsection{Output-Referred Quantization Noise}

Fig.~\ref{fig:sqnr} illustrates the improvement in the SQNR achievable by increasing the number of exponent bits for the considered distributions. The distributions that contain many large values quickly approach the SQNR ceiling because high-magnitude values dominate the global SQNR; the transition from $\mathrm{INT}$ to $\mathrm{FP_{E2}}$ offers additional codes for a modest benefit. However, further increasing exponent bits offers little benefit, as the improved fidelity of small values is not captured by global SQNR. In the uniform case, this effect is most pronounced, with half the values falling under the largest exponent; additional exponent bits refine the grid near the midpoint but offer negligible benefit to SQNR. Similarly, the max-entropy distribution maximizes SQNR by definition. Herein lies the danger of using global SQNR to specify analog hardware for narrow core distributions with rare large outliers: Even a small fraction of outliers dominates this metric and fails to provide insight into the performance on the core distribution. This effect is particularly acute in the Gaussian + outliers example, which mimics the statistics of LLM activations. Therefore, it is also instructive to consider the SQNR of the subset of samples which do not contain outliers; this is shown in Fig.~\ref{fig:sqnr}: `Gaussian + Outliers (Core)'. At $N_{\mathrm{E},x}=2$ the global SQNR is $\approx 18$, even though the core distribution containing the majority ($\approx 99\%$) of the data produces no signal because it falls below the first quantization step's rounding boundary. At $N_{\mathrm{E},x}=3$ the core distribution is resolved to within 6 dB of the ceiling, and SQNR plateaus at $N_{\mathrm{E},x}=4$.

\begin{figure}[!t]
	\centering
	\includegraphics[width=3in]{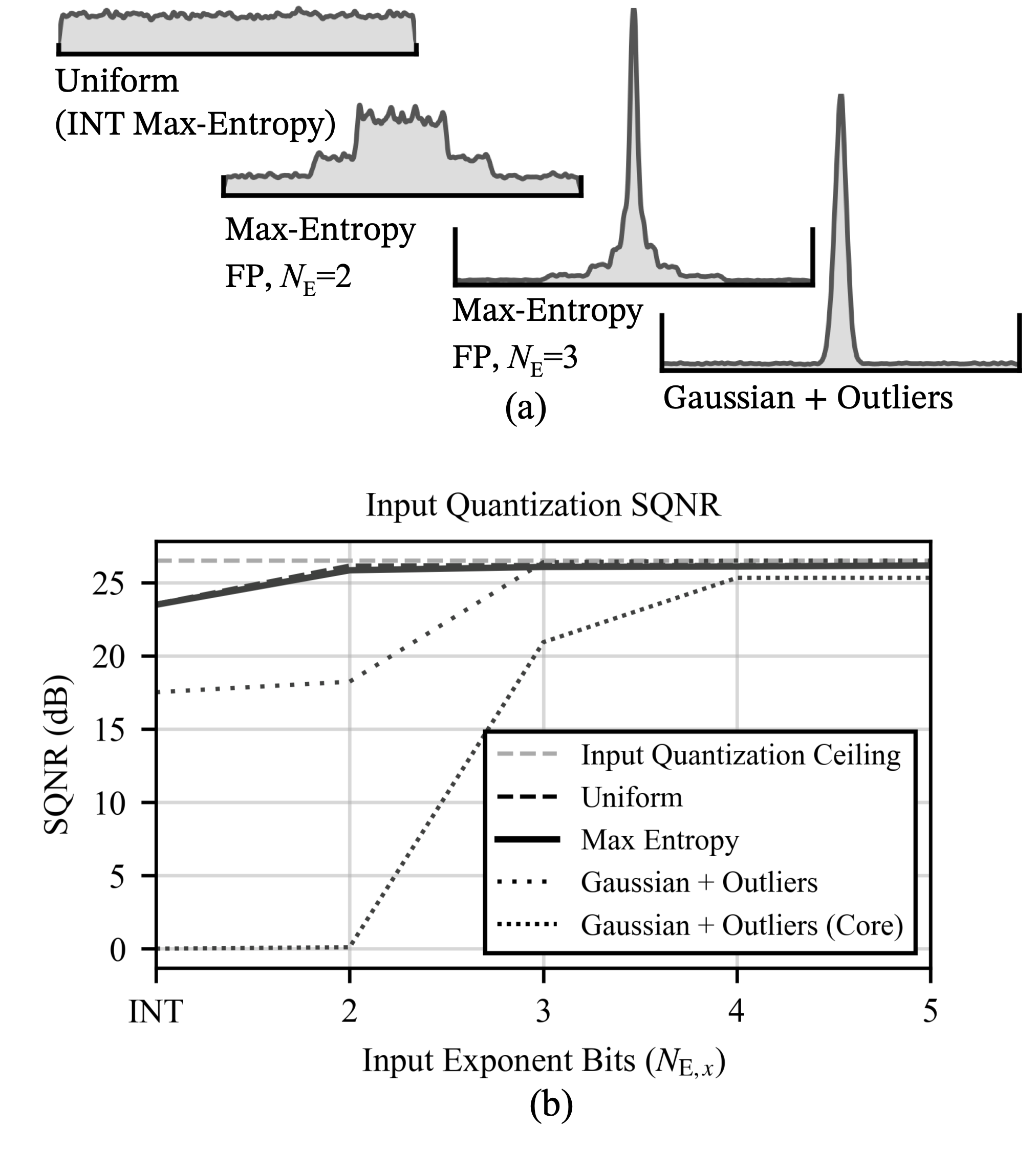}
	\caption{(a) Distributions used for evaluating CIM ADC requirements. (b) Quantization noise vs exponent bits for the illustrated distributions. $N_\mathrm{M}=2$ is used. For visual clarity, the Gaussian + outliers distribution is illustrated with an exaggerated Gaussian core width and outlier probability.}
	\label{fig:sqnr}
\end{figure}

\begin{figure}[!t]
	\centering
	\includegraphics[width=3.5in]{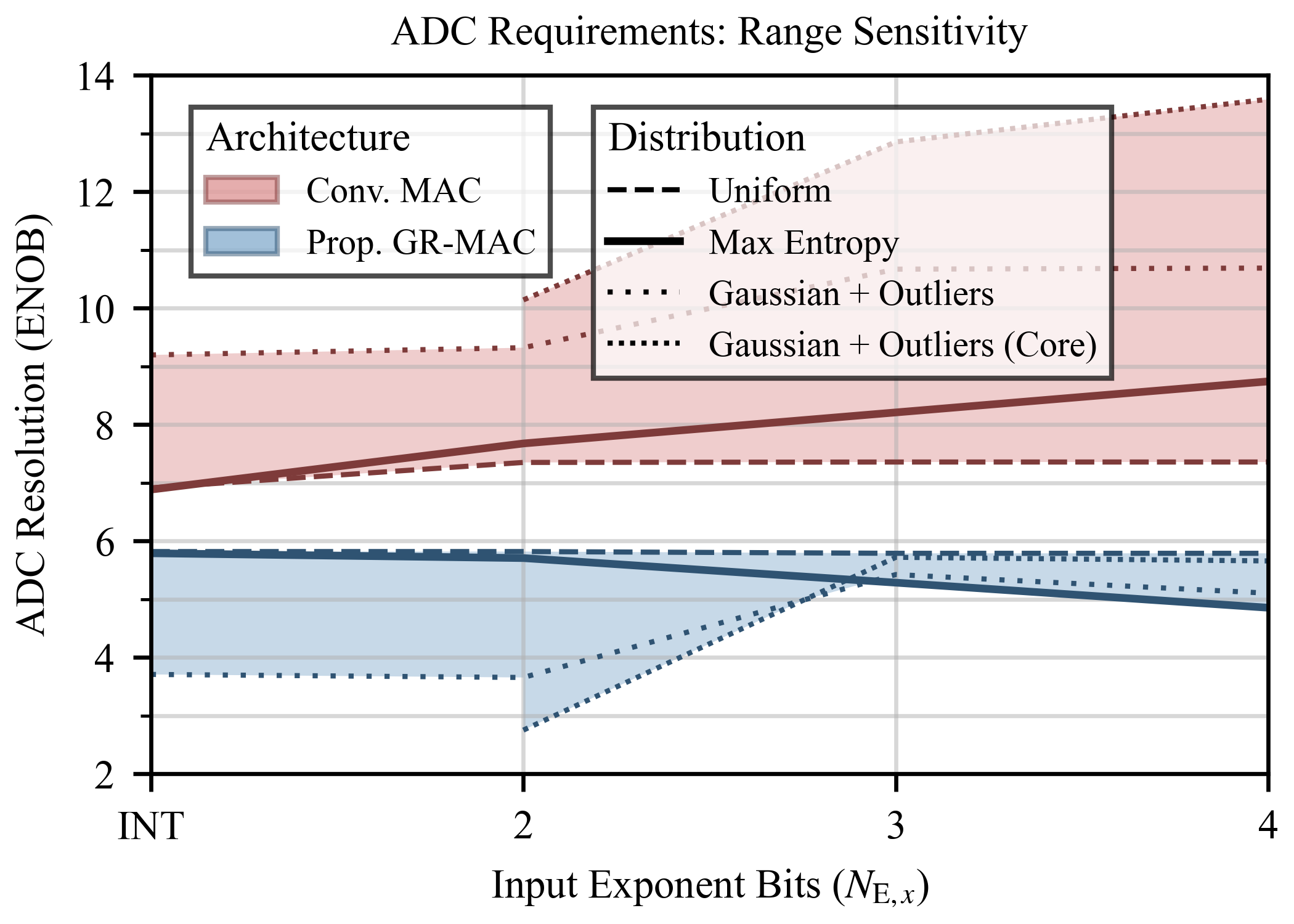}
	\caption{Required ADC resolution, $\mathrm{ENOB}$, as a function of input dynamic range, parameterized by exponent bits, $N_{\mathrm{E},x}$. The input format uses $N_{\mathrm{M},x}=2$ mantissa bits; sensitivity to input mantissa bits is minimal regarding range scaling trends, affecting ENOB by a constant offset (see Fig.~\ref{fig:anal2}). Weights are fixed to $\mathrm{FP4_{E2M1}}$ and follow a maximum-entropy distribution, chosen as an information-optimal first-order approximation of empirical weights in \cite{aciq}. The choice of weight format has a negligible effect on output-referred SQNR since only input quantization noise is considered, and for accurate results needs only be sufficiently large to avoid sparse lattice artifacts in the output. $\mathrm{ENOB}$ is specified to exceed the output-referred SQNR by 6 dB for robust conversion. To compute $\mathrm{ENOB}$, the ADC step size, $\Delta$, is related to its quantization noise using the standard formulation assuming a uniform distribution between points on the quantization grid, $P_{\mathrm{q, ADC}} = \nicefrac{\Delta^2}{12}$, and $\mathrm{SNR} \approx \mathrm{SQNR}$, giving $\mathrm{ENOB} = \log_2(\nicefrac{V_{\mathrm{FS}}}{\Delta})$. $N_\mathrm{R}=32$ rows are used.}
	\label{fig:anal1}
\end{figure}

\begin{figure}[!t]
	\centering
	\includegraphics[width=3.5in]{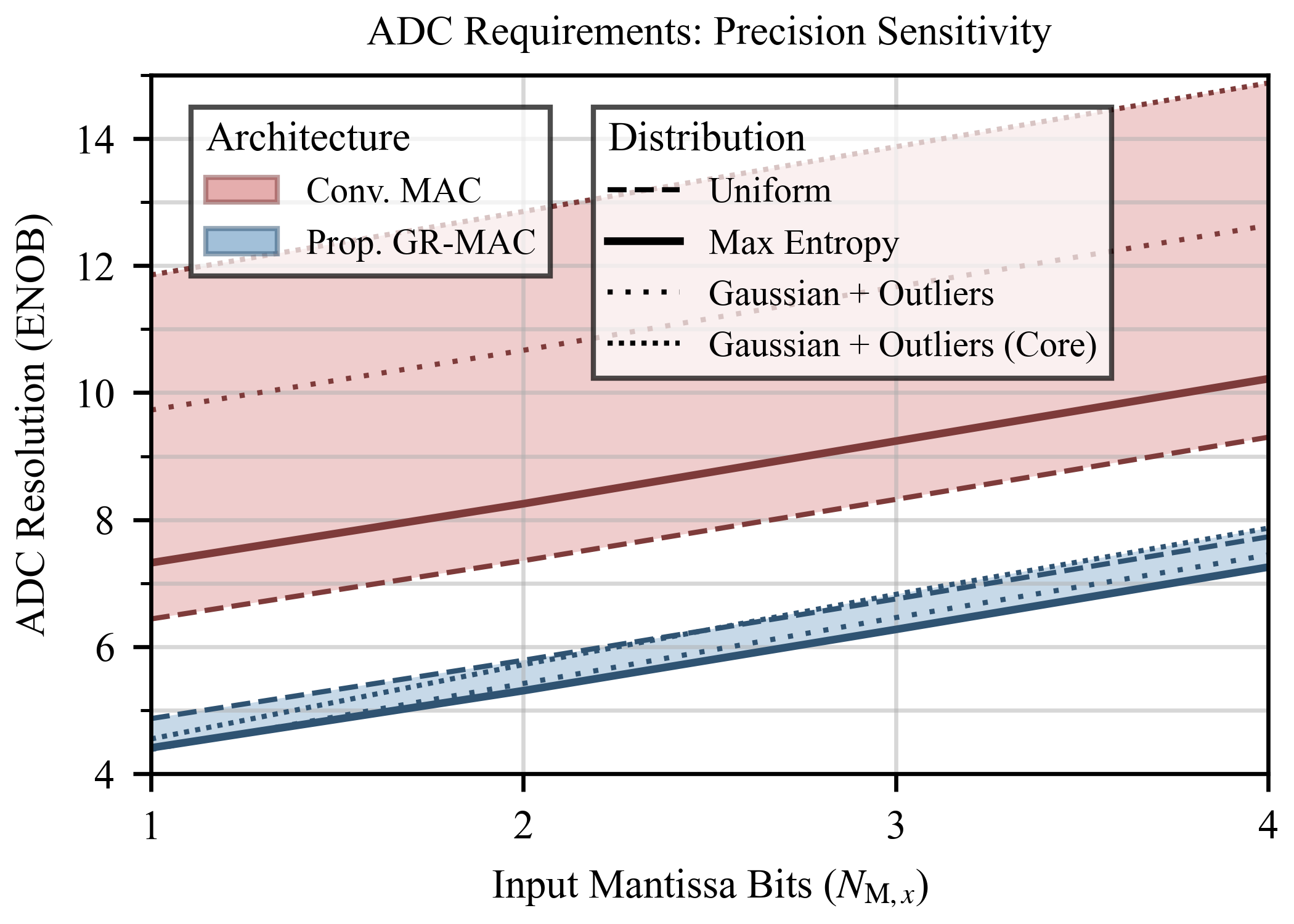}
	\caption{Required ADC resolution, $\mathrm{ENOB}$, as a function of input precision, parameterized by mantissa bits $N_{\mathrm{M},x}$. The input format uses $N_{\mathrm{E},x}=3$ to ensure that the studied distributions fall within the format's dynamic range. ENOB is computed as in in Fig.~\ref{fig:anal1}; similarly, weights are fixed to $\mathrm{FP4_{E2M1}}$, and follow a maximum-entropy distribution, and $N_\mathrm{R}=32$ rows are used.}
	\label{fig:anal2}
\end{figure}

\subsubsection{Range Sensitivity}
The required ADC resolution as a function of input dynamic range, shown in Fig.~\ref{fig:anal1}, highlights the advantage of the proposed GR-MAC. The uniform distribution lower-bounds the conventional ADC $\mathrm{ENOB}$ requirement, as it presents the most efficient utilization of the full-scale range but does not represent a practical workload, resulting in ADC underspecification. By contrast, gain-ranging introduced in the proposed architecture offers the least benefit under the uniform distribution, as the largest magnitude bins are most populated, and serves as a practical upper bound which can be used for ADC specification. The proposed upper bound demonstrates a 1.5\nbhyphen bit improvement over the conventional lower bound. Moreover, as the input dynamic range increases to capture the Gaussian + outliers core distribution ($N_{\mathrm{E},x}\ge 3$), the ADC $\mathrm{ENOB}$ improvement of the proposed GR-MAC exceeds 6 bits. Crucially, the GR-CIM ADC resolution remains below the $N_{\mathrm{cross}}\approx  10$~bits thermal-noise-regime boundary for the ADC energy model defined in the Appendix. The ADC $\mathrm{ENOB}$ requirement is suppressed for the Gaussian + outliers distribution at low $N_{\mathrm{E},x}$, due to the quantization fidelity as shown in Fig.~\ref{fig:sqnr}.

\subsubsection{Precision Sensitivity}
The required ADC resolution scales linearly with the input format's precision, shown in Fig.~\ref{fig:anal2}, as the increased input precision creates a corresponding increase in ADC resolution. The 1.5-6~bit $\mathrm{ENOB}$ advantage of the proposed architecture is observed independent of the input resolution.

\subsection{Energy}

\begin{figure*}[!t]
	\centering
	\includegraphics[width=7.16in]{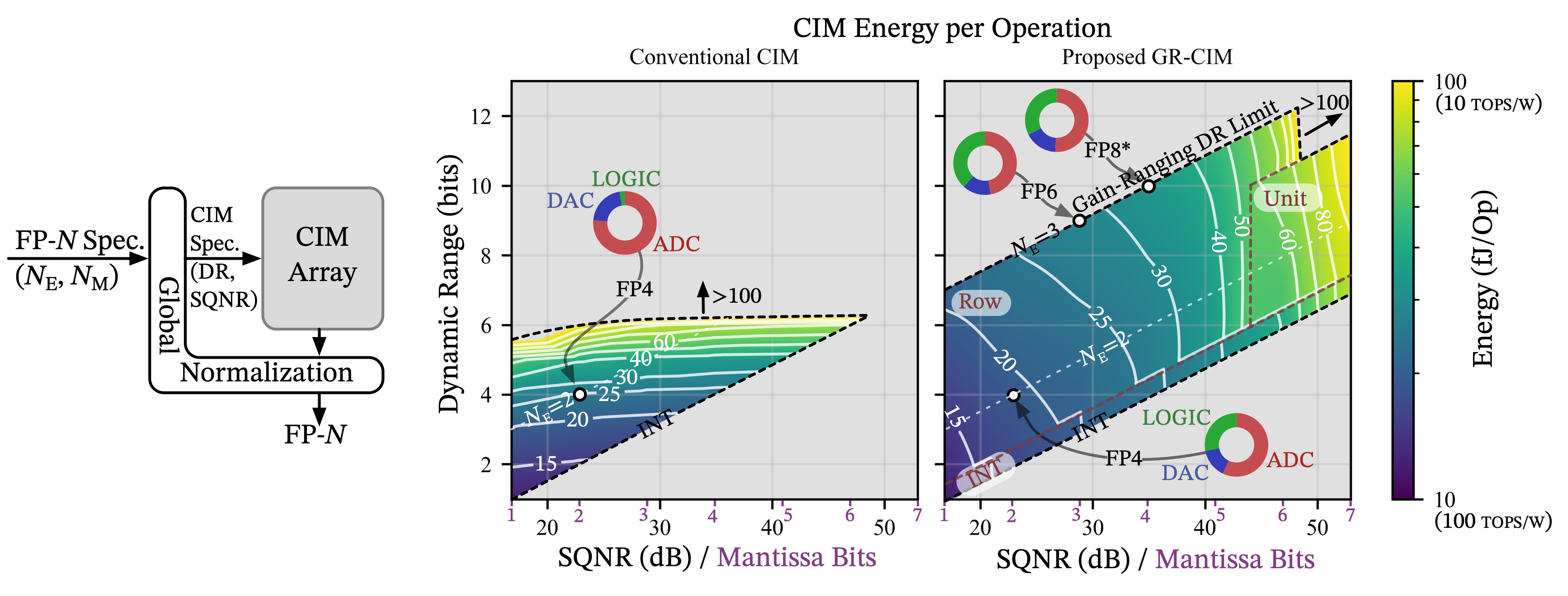}
	\caption{CIM energy per operation according to input dynamic range (DR) and precision (SQNR) spec. The GR-MAC's normalization granularity may be INT, Row, or Unit (defined in Section \ref{sec:arch}), and the energy-optimal regimes are annotated and delineated by dark-red dotted lines. Black or white dotted lines labelled $\mathrm{INT}$, $N_\mathrm{E}=2$, and $N_\mathrm{E}=3$ correspond to input formats with the specified number of exponent bits. Pie charts provide energy breakdowns for processing FP4, FP6, and FP8 inputs. $\textrm{FP4}_{\textrm{E2M1}}$ and $\textrm{FP6}_{\textrm{E3M2}}$ are shown natively without global normalization, while $\textrm{FP8}^*_{\textrm{E4M3}}$ requires global normalization to limit DR to a practical regime; only CIM array energy is included, as global normalization is independent of the CIM implementation. As the ADC specification is sensitive to the weight distribution but not the weight resolution, FP4 max-entropy weights are used, consistent with the previously established empirical observations and the information-optimality assumption. A $32\times 32$ CIM array is used. Each MAC is two operations. Mantissa bits include the implicit leading bit.}
	\label{fig:egy}
\end{figure*}

The capability of the CIM array is specified by the input format it can robustly process, parameterized by dynamic range (DR), and precision (SQNR). The energy per operation of the conventional CIM and proposed GR-CIM are modelled and compared across these dimensions. While a given SQNR corresponds directly to a minimum DR (i.e. the bitwidth of an INT format, for a uniform input), the objective of this work is to decouple these metrics to enable an increased DR without requiring a corresponding increase in SQNR. For a floating-point format, the number of mantissa bits defines the SQNR requirement, while the number of exponent bits defines the additional DR.

For a given specification, the ADC is dimensioned to robustly ($\mathrm{SNR}_{\mathrm{ADC}} \ge \mathrm{SQNR}_x + 6~\mathrm{dB}$) process a uniform input scaled to its narrowest valid bounds. This bound is defined as twice the minimum normal value and represents the smallest range which is quantized to the target SQNR. Thus, the excess DR beyond the minimum required to achieve a given SQNR manifests as a relative reduction in the input signal amplitude with respect to the full-scale range, setting the ADC's worst-case resolution bound. The DAC is dimensioned under the same conditions: Truncation is not performed as it would necessarily violate the SQNR specification. Thus, the DAC resolution increase for the conventional CIM matches the excess DR requirement, but remains constant for the proposed CIM, depending only on the SQNR requirement. The conventional CIM logic is specified to switch the capacitive dividers in each unit cell. In addition to switching the capacitor dividers, the proposed CIM provisions unit-cell exponent adders and decoders, exponent-adder trees, and multipliers for output normalization, depending on the normalization granularity, as defined in Section \ref{sec:arch}. The energy models for the ADC, DAC, and digital logic components are described in the Appendix, together with the parameters for 28 nm.

Fig.~\ref{fig:egy} presents CIM energy per operation, contrasting the scaling trends of the conventional and proposed CIMs. The valid design space is bounded along the `INT' line by the minimum DR required to achieve a given SQNR. A practical upper limit for energy of 100 fJ/Op (10 TOPS/W) is also used, which roughly coincides with the end of technology-limited ADC scaling under the parameters of this analysis. The GR-CIM's efficient design regime is further bounded by the limited dynamic range of the gain-ranging stage; it is possible to increase dynamic range past this threshold, but only by scaling in the same manner as the conventional CIM. 

The slopes of the energy contours indicate the relative cost of SQNR and DR throughout the design space: the ideal contour line is vertical, indicating that energy scales solely with precision, and allowing for increased input range without increasing energy consumption. A horizontal contour line indicates that the energy required to achieve a given range is dictated by the precision needed to achieve that range natively (along the INT line); thus, for a constant range, energy savings cannot be realized by reducing precision. The conventional CIM demonstrates this unfavourable DR-dominated scaling in Fig.~\ref{fig:egy}. By contrast, the proposed GR-CIM exhibits SQNR-dominated scaling; increasing the required precision demands a corresponding increase in ADC energy according to well-known scaling trends \cite{murmann-adc-survey, adcsurvey}, while DR may be increased at the cost of additional normalization circuitry. In addition to favourable energy scaling, gain-ranging provides a constant benefit by normalizing inputs and weights, regardless of the DR requirement.

At the 100 fJ/Op limit, the GR-CIM can robustly process inputs with 6~bits more dynamic range than the conventional CIM, while maintaining the same SQNR ($47 \mathrm{~dB}$). Zhao et al. \cite{fpcim} state that 35 dB is sufficient for most Edge-AI applications; at this level, the proposed GR-CIM achieves $4~\mathrm{bits}$ more dynamic range for the same $30~\mathrm{fJ/Op}$ energy cost.

For the common low-bit formats \cite{ocpfmt}, at $\mathrm{FP4_{E2M1}}$, the gain-ranging improves energy per operation by 23\%. The $\mathrm{FP6_{E3M2}}$ format is outside the practical range for the conventional CIM and would require global normalization, adding additional energy cost and degrading fidelity, while the GR-CIM can process it natively at $29~\mathrm{fJ/Op}$. The large dynamic range of an $\mathrm{FP8}$ format with $N_{\mathrm{E},x} \ge 4$ requires global normalization for either architecture. However, the proposed CIM extends the DR of the energy-efficient regime by $6~\mathrm{bits}$ compared to the fixed-point baseline. Consequently, when employed within a global normalization and segmentation scheme \cite{fpcim}, the GR-CIM supports a significantly wider envelope per segment, minimizing segmentation overhead.

Fig.~\ref{fig:egy} also shows energy breakdowns, illustrating how the fine-grained normalization of the GR-MAC improves ADC energy but incurs additional logic overhead. Moving forward to advanced nodes with more favourable digital scaling is expected to reduce this overhead.

\subsubsection*{ADC Energy Model Parameter Sensitivity}
As the ADC energy represents a large and often dominant portion of overall CIM energy, the choice in ADC energy parameters is highly influential on overall energy-efficiency results. The effect of varying the ADC energy parameters $k_1$ and $k_2$ is studied at the FP4 point, illustrated for both the conventional and proposed architectures. With the nominal values of $k_1$ and $k_2$ in Table \ref{tab:parameters} from \cite{sun-aord}, the proposed design shows a 23\% improvement in energy efficiency. If these factors are increased by 10\%, the proposed architecture offers a 25\% improvement in energy efficiency, whereas if they decrease by 10\%, the proposed architecture's advantage is reduced to 21\%. While the overall energy efficiency for both architectures is strongly dependent on the ADC energy, the relative advantage of the proposed architecture remains robust to perturbations to the ADC parameters.

\vspace{\baselineskip}
Overall, local normalization by adding gain-ranging to the analog MAC shows potential to significantly improve the energy and dynamic range envelope of the analog CIM.

\section{Conclusion}\label{sec:conclusion}
This work proposes adding a gain-ranging stage to the charge-based MAC to enable native analog processing of floating-point formats. Fine-grained normalization is shown to lower the ADC resolution requirement, enabling higher dynamic range inputs without global normalization or segmentation and the associated energy overhead and fidelity loss. Furthermore, an upper bound is provided for the ADC specification, decoupling it from input distribution assumptions. Energy modelling shows that GR-MAC scaling is SQNR-dominated rather than dynamic-range-dominated, enabling increased input dynamic range at minimal energy cost. At a 35 dB standard, the GR-MAC provides a 4-bit increase in the input dynamic range without increased energy consumption. The GR-MAC guides the future role of analog compute for energy-efficient processing of AI workloads. Overall, these results support a broader trend of leveraging digital scaling to improve analog performance.

\appendix
The analysis in this paper relies on models and parameters developed and validated by Sun et al. in \cite{sun-aord} against previously published analog and digital CIM implementations \cite{aord-alg-2, aord-alg-3, aord-alg-4, fpaligncim, aord-dig-6, digfpcim, intensive-cim-sparse-digital}. They are summarized in Tables \ref{tab:energy_models} and \ref{tab:parameters} and explained here.

\begin{table}[ht]
	\centering
	\begin{threeparttable}
		\caption{Energy Models for CIM Components}
		\label{tab:energy_models}
		\begin{tabular}{@{}ll@{}}
			\toprule
			\textbf{Component} & \textbf{Energy} \\ 
			\midrule
			ADC & $(k_1 \mathrm{ENOB} + k_2 4^{\mathrm{ENOB}}) V_{\mathrm{DD}}^2$ \\ \addlinespace
			DAC & $k_3 \mathrm{DAC}_{\mathrm{res}} V_{\mathrm{DD}}^2$ \\ \addlinespace
			Cell array switching & $0.5 C_{\mathrm{gate}} V_{\mathrm{DD}}^2 N_{\mathrm{SW}} N_{\mathrm{R}} N_{\mathrm{C}}$ \\ \addlinespace
			Full Adder (FA) & $6 C_{\mathrm{gate}} V_{\mathrm{DD}}^2$ \\ \addlinespace
			Adder Tree & $E_{\mathrm{FA}} \cdot \#\mathrm{FA}$ \\ \addlinespace
			$N$-bit multiplier\tnote{a} & $(1.5C_{\mathrm{gate}} V_{\mathrm{DD}}^2 + E_{\mathrm{FA}}) N^2$ \\ \addlinespace
			Binary Decoder\tnote{a} & $(0.5 N_{\mathrm{in}} + N_{\mathrm{out}} + 1) C_{\mathrm{gate}} V_{\mathrm{DD}}^2$ \\ 
			\bottomrule
		\end{tabular}
		\begin{tablenotes}[para, flushleft]
			\scriptsize
			\item[a] Derived from \cite{sun-aord} via standard logic scaling.
			\item Definitions: $\mathrm{ENOB}$: Effective ADC resolution, $\mathrm{DAC}_{\mathrm{res}}$: DAC resolution, $N_{\mathrm{SW}}$: Number of switches per cell, $N_{\mathrm{R}}, N_{\mathrm{C}}$: Number of rows and columns in the CIM array, $N_{\mathrm{in}}, N_{\mathrm{out}}$: Decoder input and output bit-width.
		\end{tablenotes}
	\end{threeparttable}
\end{table}

\begin{table}[ht]
	\centering
	\caption{Cost Model Parameters @ 0.9 V, 28 nm}
	\label{tab:parameters}
	\begin{tabular}{@{}ll@{}}
		\toprule
		\textbf{Parameter} & \textbf{Value} \\ 
		\midrule
		$C_{\mathrm{gate}}$ & $0.7~\mathrm{fF}$ \\ 
		$k_1$ & $100~\mathrm{fF}$ \\ 
		$k_2$ & $1~\mathrm{aF}$ \\ 
		$k_3$ & $50~\mathrm{fF}$ \\ 
		\bottomrule
	\end{tabular}
\end{table}

\begin{enumerate}
	\item \textit{ADC:} The ADC energy model from \cite{sun-aord} fits $k_1$ and $k_2$ according to empirical scaling limits (excluding outliers), consistent with \cite{murmann-mscim, murmann-adc-survey}. As explained in Section \ref{sec:anal}, the required ENOB is computed to ensure that the overall ADC $\mathrm{SNDR_{ADC}}\approx \mathrm{SNR_{ADC}}$ exceeds the input SQNR referred to the ADC by a safety margin (6 dB), using statistical simulations on various input conditions, rather than the closed-form solution proposed in \cite{murmann-mscim} which is limited to signal shrinkage based on full-scale uniform inputs.
	\item \textit{DAC:} The DAC energy model proposed and validated in \cite{sun-aord} models the energy cost according to $k_3$, a technology coefficient characterizing the effective switching capacitance per bit.
	\item  \textit{Logic:} 
	\begin{enumerate}
		\item \textit{Cell array switching:} This captures the total energy of bitline switching \cite{sun-aord}. Each cell contains $N_{\mathrm{SW}}$ capacitor switches, each presenting a load of two transistors to the bitline. Accordingly, the cell capacitance is normalized to $0.5C_{\mathrm{gate}}$, where $C_{\mathrm{gate}}$ is the gate capacitance of a reference four-transistor NAND2/NOR2 gate. The proposed architecture introduces a gain-ranging stage that increases $N_{\mathrm{SW}}$ by 1, as the one-hot encoded exponent control toggles once per operation.
		\item \textit{Full adder (FA) and adder tree:} The energy of a full adder (FA) is modelled as $6 C_{\mathrm{gate}} V_{\mathrm{DD}}^2$, accounting for the dominant internal node and output switching activity per addition \cite{sun-aord}. This model applies to the adder trees for digital accumulation and the unit-cell arithmetic in the unit-normalization regime.
		\item \textit{N-bit multiplier:} The energy of an N-bit multiplier is modelled as  $(1.5C_{\mathrm{gate}} V_{\mathrm{DD}}^2 + E_{\mathrm{FA}}) N^2$, reflecting the quadratic scaling of the number of AND gates and FAs required to generate and accumulate partial products.
		\item \textit{Binary Decoder:} The binary decoder controls the gain-ranging switched-capacitor array. Its energy is modelled as $(0.5 N_{\mathrm{in}} + N_{\mathrm{out}} + 1) C_{\mathrm{gate}} V_{\mathrm{DD}}^2$, accounting for $N_{\mathrm{in}}$ inverters to generate complement signals, and $N_{\mathrm{out}}$ NAND gates and inverters, with the final inverters accounting for a constant $(0.5 \cdot 2)$ as only two outputs toggle per cycle. 
	\end{enumerate}
\end{enumerate}

% BIBLIOGRAPHY
\bibliographystyle{IEEEtran}
\bibliography{cdr-labs-ai}

% AUTHORS
\begin{IEEEbiography}[{\includegraphics[width=1in,height=1.25in,clip,keepaspectratio]{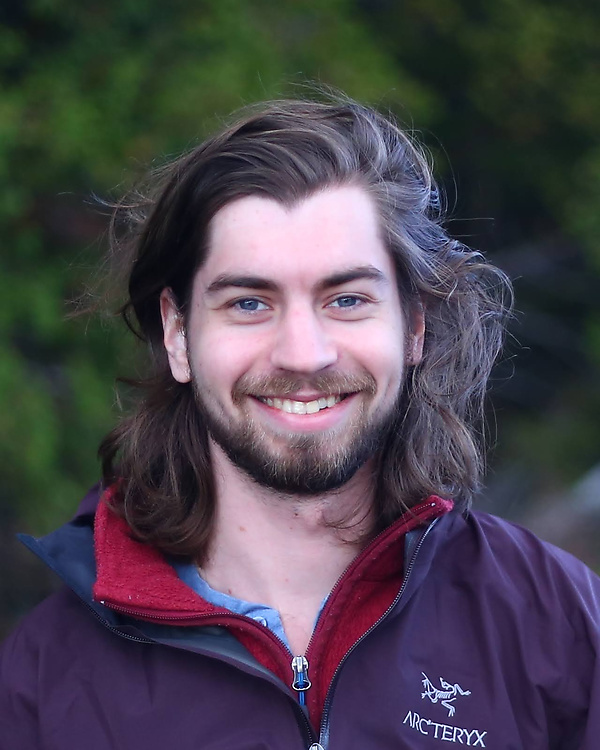}}]{Brian Rojkov} received the B.A.Sc. degree in mechatronics engineering from the University of Waterloo in 2022, where he is pursuing the Ph.D. degree in electrical and computer engineering.
\end{IEEEbiography}

\begin{IEEEbiography}[{\includegraphics[width=1in,height=1.25in,clip,keepaspectratio]{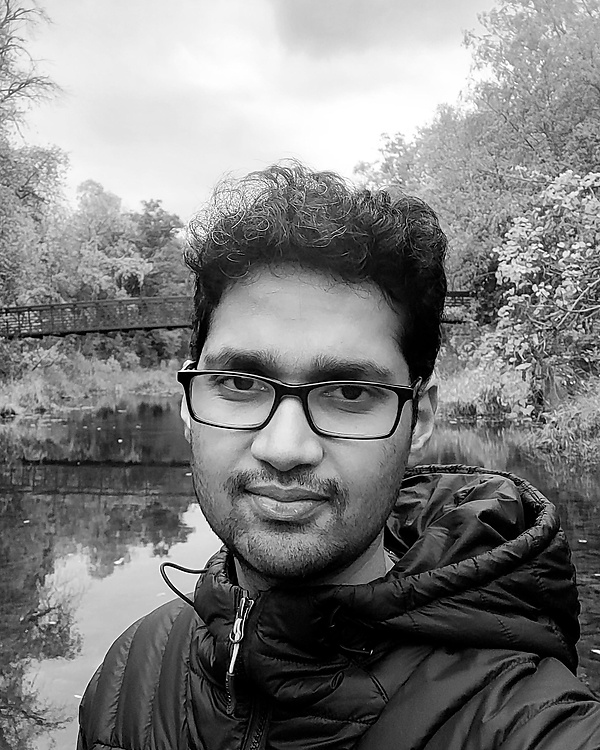}}]{Shubham Ranjan} received the Ph.D. degree in electrical and computer engineering from the University of Waterloo, Waterloo, ON, Canada, in 2025. He is currently a Postdoctoral Researcher with the Department of Electrical and Computer Engineering, University of Waterloo, Waterloo, ON, Canada. His research interests include low-power OLEDoS microdisplays, embedded memories, in-memory computing, flexible electronics, and thin-film transistor (TFT)-based circuits and systems.
\end{IEEEbiography}

\begin{IEEEbiography}[{\includegraphics[width=1in,height=1.25in,clip,keepaspectratio]{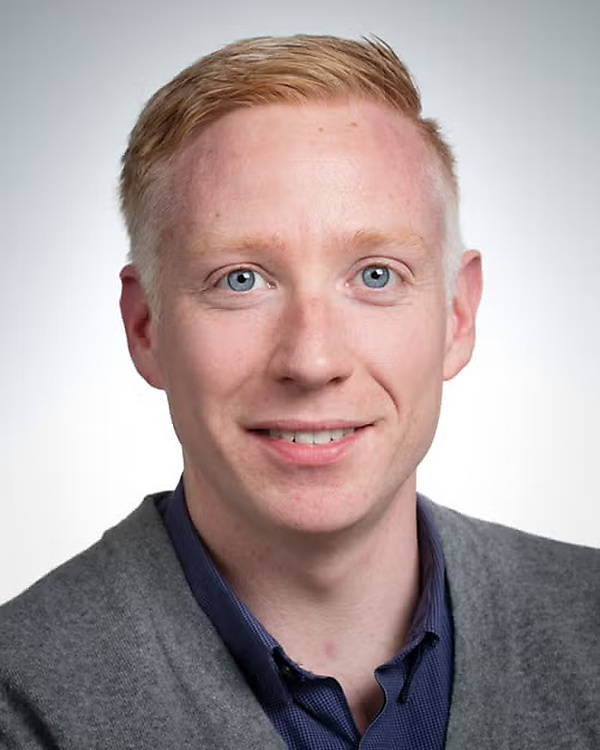}}]{Derek Wright} received the B.A.Sc. and M.A.Sc. degrees in electrical and computer engineering from the University of Waterloo in 2003 and 2005, respectively, and the Ph.D. degree in collaborative electrical and biomedical engineering from the University of Toronto in 2010. He is currently an Associate Professor with the Electrical and Computer Engineering Department at the University of Waterloo, where he also serves as Director of Mechatronics Engineering. His research interests include low-power digital circuit design, multidomain modeling and simulation, biomedical devices, and neuromorphic circuits and systems.
\end{IEEEbiography}

\begin{IEEEbiography}[{\includegraphics[width=1in,height=1.25in,clip,keepaspectratio]{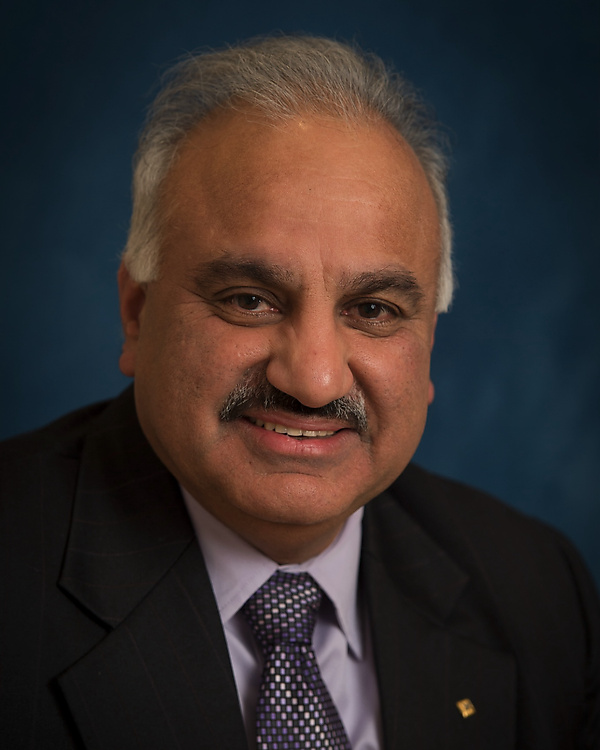}}]{Manoj Sachdev} is a professor of Electrical and Computer Engineering at the University of Waterloo, Canada since 1998. His current research interests include low-power and high-performance digital circuit design, mixed-signal circuit design, test and manufacturing issues of integrated circuits. He has contributed to five books, two chapters, and has co-authored over 300 technical articles in conferences and journals. He holds more than 50 granted and pending US patents on various aspects of VLSI circuit design, reliability, and test. He, his students, and his colleagues have received several international research awards. In 1997, at the IEEE European Design and Test Conference, he received the best paper award. In 1998, he was a co-recipient of the honorable mentioned award in the IEEE International Test Conference. He received the best panel award in 2004 IEEE VLSI Test Symposium. In 2011, he was a co-recipient of the best paper award in IEEE International Symposium on Quality Electronics Design. In 2015, he was a co-recipient of the best poster award at IEEE Custom Integrated Circuits Conference.
	
Professor Sachdev received his B.E. degree (with Honors) in Electronics and Communication Engineering from IIT Roorkee (India), and Ph.D. from Brunel University (UK). He was with Semiconductor Complex Limited, Chandigarh (India) from 1984 till 1989 where he designed CMOS Integrated Circuits. From 1989 till 1992, he worked in the ASIC division of SGS-Thomson at Agrate (Italy). In 1992, he joined Philips Research Laboratories, Eindhoven (The Netherlands), where he researched on various aspects of VLSI testing and manufacturing. 

Professor Sachdev is an IEEE Fellow, Fellow of Engineering Institute of Canada, and Fellow of Canadian Academy of Engineering. He has served on committees of several IEEE circuit conferences such as CICC, DATE, and VLSI Design. 	
\end{IEEEbiography}
\vfill

\end{document}